\documentclass[onecolumn,floats,showpacs,prd,amssymb,floatfix,nofootinbib,balancelastpage,superscriptaddress,amsmath,showkeys]{revtex4}
\usepackage{epsfig}

\def\be{\begin{equation}}
\def\ee{\end{equation}}
\newcommand{\ds}{{\sffamily DarkSUSY}}
\newcommand{\Dpp}{D_{pp}}
\newcommand{\Dxx}{D_{xx}}
\newcommand{\ddp}{{\partial\over\partial p}}
\def\msun{M_{\odot}{\ }}

\begin{document}

\title{Testing the Dark Matter Interpretation of the PAMELA Excess through Measurements of the Galactic Diffuse Emission}

\author{Marco Regis}
\email{Marco.Regis@uct.ac.za} 

\affiliation{Cosmology and Gravity Group, Department of Mathematics and Applied
Mathematics, University of Cape Town, Rondebosch 7701, South Africa}

\author{Piero Ullio}
\email{ullio@sissa.it}

\affiliation{SISSA,
              Via Beirut 2-4, I-34014 Trieste, Italy and\\
              Istituto Nazionale di Fisica Nucleare,
              Sezione di Trieste, I-34014 Trieste, Italy}


\begin{abstract}

We propose to test the dark matter (DM) interpretation of the positron  
excess observed by the PAMELA cosmic-ray (CR) detector through the identification of a  
Galactic diffuse gamma-ray component  associated to DM-induced prompt and radiative emission.
The goal is to present an analysis based on minimal sets of assumptions and extrapolations with 
respect to locally testable or measurable quantities. We
discuss the differences between the spatial and spectral features for  
the DM-induced components (with an extended, possibly spherical, source function)
and those for the standard CR contribution
(with sources confined within the stellar disc), and propose to focus  
on intermediate and large latitudes. We address the dependence of  the signal to background  
ratio on the model adopted to describe the propagation of charged CRs in the Galaxy, 
and find that, in general, the DM-induced signal can be detected by the Fermi  
Gamma-ray Space Telescope at energies above 100 GeV. An observational result in agreement with  
the prediction from standard CR components only, would imply very strong constraints on  
the DM interpretation of the PAMELA excess. On the other hand, if an excess in the diffuse  
emission above 100~GeV is identified, the angular profile for such emission would allow for a  
clean disentanglement between the DM interpretation and astrophysical explanations proposed  
for the PAMELA excess.
We also compare to the radiative diffuse emission at lower frequencies, sketching in particular
the detection prospects at infrared frequencies with the Planck satellite.
\end{abstract}




\maketitle

\section{Introduction}

Recent measurements of the positron fraction in cosmic rays (CRs) up to 100~GeV by the PAMELA 
experiment~\cite{Adriani:2008zr} (and of the sum of the electron plus positron fluxes by 
ATIC~\cite{:2008zzr} and PPB-BETS~\cite{Torii:2008xu}) have triggered a lot of interest on the possibility that there may be a dominant  
contribution to these terms from the annihilation or decay of dark matter (DM) particles, 
assumed to compose the dark halo of the Milky Way.  Indeed, the sharp raise in the positron 
fraction detected by PAMELA above 10~GeV, confirming and super-exceeding 
previous hints of an anomaly~\cite{Barwick:1997ig,Alcaraz:2000bf,Aguilar:2007yf}
is a feature which cannot be accommodated within the "standard picture", with primary electrons accelerated 
in supernova remnants (SNRs) and secondary positrons produced mainly from the interaction of 
primary cosmic-rays with the interstellar medium (ISM) during propagation.
It is instead suggestive of an extra primary source of positrons.
Pulsars are well-motivated candidates for this role, see, 
e.g.,~\cite{Aharonian,Hooper:2008kg,Profumo:2008ms}. Another interesting possibility is 
the recent proposal that positrons (and electrons) are secondary products of hadronic 
interaction within the cosmic ray sources~\cite{Blasi:2009hv}.

A contribution to the positron flux from dark matter was predicted 
long ago, see, e.g.~\cite{Rudaz:1987ry,Ellis:1988qp,Kamionkowski:1990ty}; depending on the
dark matter particle mass and on the shape of the electron/positron yield from annihilation or 
decay, the excess can be accommodated, see, e.g.,~\cite{Cirelli:2008pk}.
A few puzzling issues arise however in connection to the dark matter interpretations: the required 
positron injection rate is fairly large, and, supposing the signal is due to annihilating particles,
this points to pair annihilation rates much larger than the typical rates for thermal relic weakly 
interacting massive particles (WIMPs) and/or to substructure enhancements which should be unrealistically large. Also, it is to some extent surprising that a dark matter signal emerges 
clearly in the positron flux and it is still hiding in other indirect detection channels:  for typical 
particle physics dark matter candidates (where the "typical" here stands for the absence of 
a mechanism forcing the coupling of the dark matter particle to light leptons only) signals to 
background ratios tend to be larger for the antiproton flux (and even more, in perspective, 
for the antideuteron flux), and for the gamma-ray component produced via neutral pion decays. 
These features have guided recent work on both proposing new dark matter candidates,
as well as setting constraints on the range of those that had been previously proposed.

The issue of whether the standard cosmic ray picture could give a fair description  
of the electron/positron populations in the Galaxy and of whether the DM component could give a sizable contribution had actually been risen a few times even before the 
recent refined measurements of their local fluxes. Indeed, although there is a wide consensus in associating the observed multi-wavelength diffuse emission in the Galaxy to CR interactions with the ISM and magnetic field, CR spectra and propagation parameters tuned to describe the local picture can lead to an underestimation of the flux in some regions of the sky. 
High energy electrons emit radio to microwave photons via synchrotron radiation on the Galactic magnetic fields and an excess in the innermost region of the Galaxy has been claimed based on WMAP data (the so-called "WMAP haze") when scaling the measured 
synchrotron radio intensities to the microwave 
band~\cite{Finkbeiner:2004us,Hooper:2007kb,Dobler:2007wv,Cumberbatch:2009ji}.
Equally important is the Inverse Compton (IC) radiation component of high energy electrons on the starlight and microwave photon backgrounds. 
One of the models put forward to 
address the excess above 1~GeV reported by EGRET in the observations of the diffuse gamma-ray emission of the Galaxy~\cite{Hunger:1997we} considered the possibility 
of a dominant contribution from IC on starlight of an electron population in the Galaxy, on 
average, much larger than what is locally measured~\cite{Strong:2004de} ("optimized" Galprop model).
Recent preliminary data~\cite{FERMI:preliminary} from the FERMI gamma-ray telescope (formerly GLAST), however,
do not seem to confirm the EGRET excess, with the level of the diffuse flux at high
energy ($E\leq 10$ GeV) possibly consistent with a standard cosmic-ray setup and pointing to an instrumental 
effect in EGRET.
Other excesses, which are not commonly connected to a DM interpretation, has been reported in the X-ray emission at the Galactic Ridge~\cite{Porter:2008ve} and in the TeV $\gamma$-ray emission measured by Milagro~\cite{Abdo:2008if}.
The first relies on a quite complicate region, where simplified assumptions on the primary injection spectra and on the ISM properties can break down. The contribution of unresolved source in the Milagro data can be large~\cite{Casanova:2007cf}.
Moreover, we are mostly interested in a fair description of the local region, which involve only the diffuse emission at intermediate and high latitudes and longitudes. 

Most attempts to insert the dark matter interpretation of the PAMELA data into a more global 
picture have involved either other species such as gamma-ray and antiproton yields, or other
dark matter environments (e.g., the central region of the Galaxy, or Galactic satellites, instead
of the local dark matter population which, in case the dark matter interpretation holds, would 
be most likely responsible for the measured positron flux). These kind of comparisons are 
inevitably model dependent. Comparing yields is certainly very powerful when having a 
reference particle physics model in mind; on the other hand, as already mentioned,  one may
just tune the dark matter candidate in such way that only light leptons are produced as 
detectable yields. Radiative components are instead unavoidably associated to 
electron/positron yields: IC emission of a 100 GeV to 1 TeV electron on $1~\mu$m starlight  photons
gives gamma-rays with energies peaked in about the range 50 GeV to 5 TeV; 
the associated synchrotron 
emission on a~1~$\mu$G magnetic field is peaked between 50 to 5000~GHz (scaling linearly
with the magnetic field). Having normalized the 
electron/positron yield to the locally observed flux, the extrapolation for the radiative emission in 
the local portion of the Galaxy and its neighborhood is fairly solid, since although 
it is true that there are uncertainties in the astrophysical model involved, such as the parameters 
in the cosmic-ray propagation model and the level of the stellar radiation and magnetic fields, 
these need to be in turn readjusted to the local measurements (regarding the local
electron flux as well as from ratios of secondaries to primaries in cosmic rays). Extrapolations to,
e.g., the central region of the Galaxy to compare, e.g., with the WMAP haze, are much more 
uncertain since they rely on the extrapolations for the astrophysical parameters (with 
various kinds of degeneracies which can be involved) as well as for the level of dark matter yield
(in connection, e.g., with the dark matter density profile).

We will study the dark matter contribution to Galactic radiation seed at intermediate to large 
latitudes. The emission in such portion of the sky can actually probe the local environment and this test of the DM interpretation of the PAMELA excess can be thus considered self-consistent.

We will analyze the dependence of the predicted signal on the uncertainties in the
cosmic-ray propagation model. We will also sketch the differences with respect to the picture
in which the source(s) accounting for the PAMELA positron fraction is confined to the stellar disc
rather than being spread to on a much larger vertical extension such as in case of the dark 
matter source. The reference to test experimentally our proposal will be the FERMI gamma-ray 
telescope (and, to a minor extent, the PLANCK satellite~\cite{PLANCK}).

Recent analyses of the diffuse Galactic emission induced by DM models in light of the PAMELA and ATIC data includes Refs.~\cite{Cholis:2008wq,Zhang:2008tb,Borriello:2009fa,Barger:2009yt,Kawasaki:2009nr,Cirelli:2009vg}.

The paper is organized as follows. In Section~\ref{sec:CRprop}, we present the description of the particle propagation in the Galaxy. In Section~\ref{sec:source}, we discuss the spectral and spatial distribution of the CR and DM sources.
In Section~\ref{sec:signal}, we present the results for the final $e^+/e^-$ distribution and for the multi-wavelength diffuse emission induced by CRs and DM. Section~\ref{sec:conc} concludes.

\section{Cosmic-ray propagation in the Galaxy}
\label{sec:CRprop}
We adopt the description of cosmic-rays as particles propagating in a determinate environment (i.e., disregarding the effects induced on the ISM by the interaction with CRs). 
The CR propagation equation for a particle species $i$ can be written in the form~\cite{Berezinskii:1990}:

\be
{\partial n_i (\vec r,p,t) \over \partial t} = 
    \vec\nabla \cdot ( \Dxx\vec\nabla n_i - \vec v_c \,n_i )
   + \ddp\, p^2 \Dpp \ddp\, {1\over p^2}\, n_i                  
   - {\partial\over\partial p} \left[\dot{p}\,n_i
   - {p\over 3} \, (\vec\nabla \cdot \vec v_c )\,n_i\right]+q(\vec r, p, t)+\frac{n_i}{\tau_f}+\frac{n_i}{\tau_r}
\label{Eq:transport}
\ee
where $n_i$ is the number density per particle momentum ($n_i(p) dp=N_i(E) dE$, with $N_i(E)dE$ being the number density in the energy interval $(E,E+dE)$), $q$ is the source term, $\Dxx$ is the spatial diffusion coefficient along the regular magnetic field lines, $\vec v_c$ is the velocity of the Galactic wind, $\Dpp$ is the coefficient of the diffusion in momentum space, $\dot{p}$ is the momentum loss rate, and $\tau_f$ and $\tau_r$ are the time scales for fragmentation loss and radioactive decay, respectively.

The transport equation is solved numerically and assuming a cylindrical symmetry, with halo boundaries at disc radius $R=20$ kpc and half-thickness $z_h$ as described below. We exploited a modified version of the GALPROP code~\cite{Strong:1998pw}. The main modifications consist in introducing by input the spatial and spectral profiles of the DM source (computed within the \ds package~\cite{Gondolo:2004sc}), and in including the possibility of a spatially varying diffusion coefficient.

In the following, we mainly consider one-zone models with isotropic diffusion, which can be regarded as the most extensively tested models of the recent past (see, e.g., Ref.~\cite{Strong:2007nh} for a review).

Our approach is to perform self-consistent tests in the local region and the parameters in Eq.~\ref{Eq:transport} are chosen to strictly reproduce the local directly-observed spectra of nuclei and electrons.

The goal of the paper is to study the possibility of disentangling the diffuse signals originated from two different sources, CRs and DM, having different spatial distributions.
The CR injection source is confined to the Galactic plane, while the DM profile has a spherical shape. The region with intermediate and large $z$ is thus the best target for the analysis.
The propagation reshuffles the distribution of the two populations of electrons (and thus IC and bremsstrahlung signals), and the $\gamma$-ray signal associated to the decays of CR pions. 
The scaling of the signal along the $z$-direction is affected by almost any quantity entering in the transport equation, such as the description of the diffusion, the wind velocity, the magnetic field structure, and the interstellar radiation field (ISRF) distribution. Moreover, it is dramatically sensitive to the height of the propagation halo, namely, to the boundary condition along the $z$-axis.

We are not interested in performing a full scan of the propagation parameters space and estimating the corresponding uncertainties in the CR spectra (see, e.g., Refs.~\cite{Maurin:2001sj,Simet:2009ne}); rather, we want to investigate how different scalings along the $z$-direction due to different propagation models can affect the predictions for the signal to background ratio. 
We consider six benchmark scenarios of propagation and injection spectra, which are summarized in Table 1. In the following, we motivate our selection.

{\small
\begin{table}[t]
\begin{center}
\begin{tabular}{|c|c|c|c|c|c|c|c|c|c|}
\hline
$\,\,\,\,\,\,\,\,\,\,\,$&$z_h$&$D_0$&$\alpha$&$v_a$&$\beta_{inj,nuc}$&$\beta_{inj,e}$& $dv_c/dz$&$\chi^2_{red}$&color\\
$\,\,\,\,\,\,\,\,\,\,\,$&kpc&$10^{28}\,{\rm cm^2s^{-1}}$& & km/s& & & km/s kpc$^{-1}$ &(d.f.=19)&coding
\tabularnewline
\hline
\hline
B0&4&3.3&1/3&35 &1.85/2.36&2.0/2.35&0&$0.67$&blue
\tabularnewline
\hline
\hline
B1&1&0.81&1/3&35 &1.65/2.36&2.0/2.35&0&$0.77$&green
\tabularnewline
\hline
\hline
B2&10&6.1&1/3&35 &1.85/2.36&2.0/2.35&0&$0.74$&red
\tabularnewline
\hline
\hline
B3&4&3.25&1/3&45 &1.85/2.36&2.0/2.35&10&$0.84$&orange
\tabularnewline
\hline
\hline
B4&4&1.68&1/2&22 &2.4/2.2&2.35/2.35&0&$0.86$&cyan
\tabularnewline
\hline
\hline
B5&10&$2.8\cdot e^{|z|/z_s}$&1/3&35&1.85/2.36&2.0/2.35&0&$0.66$&magenta
\tabularnewline
\hline
\end{tabular}
\end{center}
\caption{Benchmark models of propagation. The spectral index break for protons and electrons is at 9 GeV in the cases with Kolmogorov diffusion, and at 40 and 9 GeV, respectively, in the Kraichnan case. The scale of diffusion in the model B5 is taken to be $z_s=4$ kpc.}
\label{tabBM}
\end{table}

}

\begin{figure}[t]
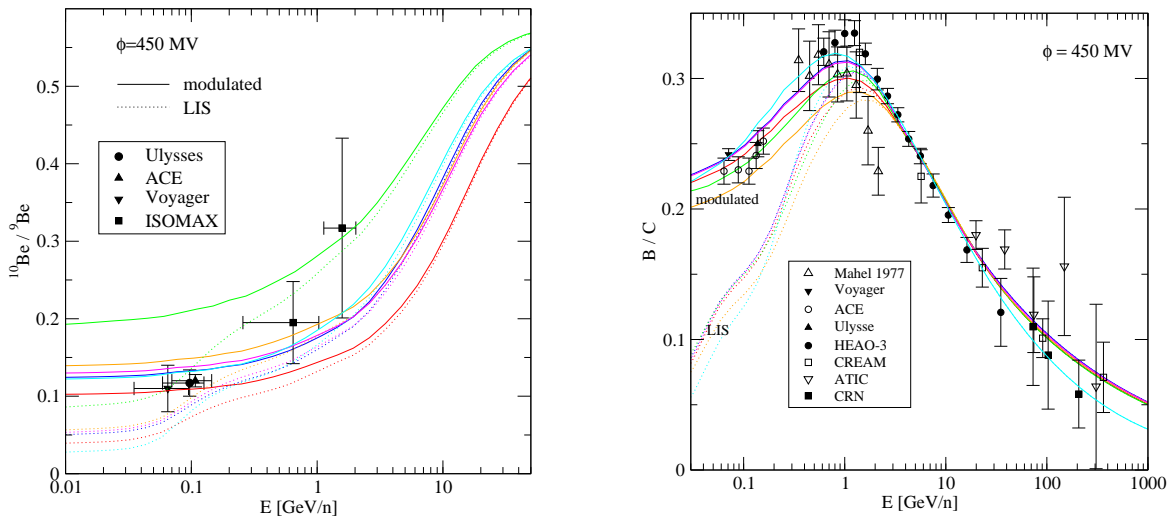

 \begin{minipage}[htb]{7cm}
   \centering
   \includegraphics[width=7.cm]{fig1a.eps}
 \end{minipage}
 \ \hspace{10mm} \
 \begin{minipage}[htb]{7cm}
   \centering
   \includegraphics[width=7.cm]{fig1b.eps}
 \end{minipage}
    \caption{Local $^{10}Be/^9Be$ ({\it Left Panel}) and B/C ({\it Right Panel}) ratios. Data are compared to the benchmark models for propagation B0 (blue), B1 (green), B2 (red), B3 (orange), B4 (cyan), and B5 (magenta). For details on the models see Table 1 and text.}
\label{fig:CRratio}
 \end{figure}

{\bf Halo height:}
In addition to the "conventional" model having $z_h=$4 kpc (named B0), we consider two models of propagation in which the halo height has been set to $z_h$=1 kpc (model B1) and $z_h$=10 kpc (model B2).
The strongest constraints on the halo height is given by the "radioactive clocks", namely, unstable secondaries. Indeed, the ratio between stable and decaying isotopes is sensitive to the CR confinement time, which is in turn related to the halo height (and the diffusion coefficient). At present, the most precise measurements is the ratio $^{10}$Be/$^9$Be, with the unstable $^{10}$Be decaying in $10^6$ years.
In Fig.~\ref{fig:CRratio}a, we show the local interstellar spectra (LIS) and the modulated spectra of the $^{10}$Be/$^9$Be ratio. The solar modulation is computed in the force field approximation~\cite{Gleeson:1968}.
As expected, $z_h=$4 kpc seems to be preferred by data. The model B2 is fully consistent with data at low energy (which are the most reliable), while it predicts a slightly smaller ratio than the two points at $\sim 1$ GeV (which, however, have very large error bars).
The model B1 shows, on the other hand, friction with data. 
Indeed, in the case with $z_h$=1 kpc, any model for diffusion and convection that can fit the $^{10}$Be/$^9$Be data dramatically increases the time spent by particles at low energy in the system, in order to let the $^{10}$Be decay. This leads to an overproduction of secondary, at odds, e.g., with the B/C data at low energy. 
On the other hand, the properties of the local environment are highly relevant for radioactive species, since they typically travel a short distance before decaying. The presence of a Local Bubble surrounding the Sun can significantly increase the amount of decays (lowering the ratio)~\cite{Donato:2001eq}. A precise estimate of this effect, however, would require a detailed description of the local environment (probably, a collection of clouds) and this subject has not been fully addressed yet~\cite{Strong:2007nh}. For this reason, we consider the model with $z_h$=1 kpc as disfavoured, but not definitively ruled out, and we do not disregard it.

{\bf Convection:}
CR data at energies below the "knee" are in fair agreement with the description of the CR transport in the Galaxy through a diffusive model, with just the relevance of convection in some regions under debate.
Galactic winds are driven by active galactic nuclei or starbursts~\cite{Veilleux:2005ia}. Although the Milky Way have a star-formation rate significantly lower than starburst galaxies, the diffuse soft X-ray emission in the inner region of the Galaxy can be well explained by a kpc-scale wind~\cite{Everett:2008}. This has motivated a renewed interest in model of cosmic-ray transport in which the convection plays a significant role~\cite{Breitschwerdt:2008}. 
The behaviour of the ratio of secondary to primary abundances disfavours a very high wind speed on the disc~\cite{Strong:1998pw}. Moreover, in the outer region of the Milky-Way, there are no direct observational evidence of a gas-loss in a wind as for the inner region.
On the other hand, it has been shown that models with a significant convection but an anisotropic diffusion~\cite{Breitschwerdt:2002vs}, or with two-zone of propagation~\cite{Ptuskin:1997} (a purely diffusive zone close to the disc and the diffusive/convective zone outside), or with a wind speed increasing with distance from the disc~\cite{Strong:1998pw}, are consistent with CR data.

In the CR models we are going to consider (i.e., one-zone of propagation and homogeneous diffusion), if the convection takes over at low energies, the secondary to primary ratio generally flattens, leading to conflict with data. The convection velocity $v_c$ is thus very constrained on the plane, and we take $v_c(z=0)=0$.
Models for thermally and cosmic-ray driven winds predict speeds increasing with the distance from the disc, with a linear scaling at intermediate $z$~\cite{Zirakashvili:1996}. Radioactive isotopes and B/C data lead to the constraint: $dv_c/dz \lesssim 10$ km s$^{-1}$ kpc$^{-1}$~\cite{Strong:1998pw}.
We will analyze the effect of convection on the diffuse emission in the benchmark model B3, considering $v_c =10\cdot(z$/1 kpc) km s$^{-1}$.
In such model, electrons at high latitudes (and low energy) can stream away due to the wind transport, reducing the related IC and bremsstrahlung signals.

{\bf Spatial diffusion and Reacceleration:}
The scattering of cosmic-ray particles on random hydromagnetic waves leads to diffusion and stochastic acceleration of the particles.
The process of diffusion has a resonant character and it is mainly driven by the energy density associated to the random component of the magnetic field at the the resonant wave number of the scattering, $k_{res}=1/r_g$, where $r_g$ is the gyro-radius of the particle. In the quasi-linear approximation of turbulence, the perpendicular diffusion is subdominant, and the diffusion takes place mainly along the magnetic field lines, with the the form of the diffusion coefficient~\cite{Berezinskii:1990}: $D_{xx}(r,p)=1/3 r_g v_p (\delta B(k_{res}) /B)^{-2}$, where $v_p$ is the particle velocity and $\delta B$ is the amplitude of the random magnetic field. 
Combining data on fluctuations of the thermal electron density, interstellar cloud density, and interstellar magnetic field~\cite{Armstrong:1995zc}, the spectrum of the energy density of the interstellar turbulence in the nearby region ($\lesssim 1$ kpc) of the Galaxy is found to be well described by a power law, $ k^{-2+\alpha}$. The data are consistent with a Kolmogorov-like spectrum ($\alpha=1/3$) over a large range ($10^8$ to $10^{20}$ cm) of wavelength, and the inferred outer scale of turbulence is L$\sim$100 pc.
Strong fluctuations on such large scale tend to isotropize the CR distribution.
The diffusion coefficient can be then described by the form: $D_{xx}\simeq \beta_p D_0\,R_{GV}^{\alpha}$ with $R$ being the rigidity of the particle, $\beta_p=v_p/c$ and typical value for the diffusion coefficient in the Galaxy are $D_0 \simeq 10^{27} - 10^{30} \rm{cm}^2\rm{s}^{-1}$~\cite{Strong:2007nh}.
Homogeneous diffusion can be regarded as a good approximation for the analysis of the nearby region. The picture at, e.g., large $z$ is, on the other hand, more uncertain and spatially varying diffusion coefficient can quite significantly affect the predictions for the $\gamma$-ray signals in such region of the sky~\cite{Evoli:2008dv}, while local spectra were not highly modified, being spallation mainly confined in the Galactic disc. 
We consider such possibility, by describing the diffusion coefficient as $\tilde D_{xx}=D_{xx} \exp(z/z_s)$, with $z_s=4$ kpc, in the benchmark model B5.

The stochastic reacceleration process refers to a diffusion in momentum which can be described by the coefficient $D_{pp}=p^2 v_a^2/(9\,D_{xx})$, where $v_a$ is the Alfv$\acute{e}$n velocity associated to the propagation of the hydromagnetic waves~~\cite{Berezinskii:1990}.
The expressions considered for $D_{xx}$ and $D_{pp}$ can describe the isotropic distribution at large scale of the Alfv$\acute{e}$n and fast magnetosonic waves (interactions between particles and slow magnetosonic waves are usually small).
Numerical and analytical treatments of the Alfv$\acute{e}$n waves cascades typically leads to a Kolmogorov spectrum for the interstellar turbulence~\cite{Ptuskin:2005ax}; this case, due to the fast cascade rate, is not significantly affected by damping on cosmic rays and we disregard this effect.
Typical values for the Alfv$\acute{e}$n speed in the Galaxy are $v_a\sim 10-30$ km/s.

In addition to the Kolmogorov case, we consider a Kraichnan spectrum of interstellar turbulence, i.e., $\alpha = 0.5$. In this case, damping on cosmic rays can significantly affect the reacceleration mechanism and we include it following the line of~\cite{Ptuskin:2005ax} (taking the damping constant $g = 0.06$ and the maximum free path length of $15$ pc). This model of propagation is labeled B4 in Table 1.  

The diffusion coefficient $D_0$ and Alfv$\acute{e}$n velocity $v_a$ are tuned in order to reproduce the B/C ratio in all the benchmark models. Values are reported in Table 1.
Spectra of the B/C ratio are shown in Fig.~\ref{fig:CRratio}b.
Note that all the benchmark models are satisfactory in fitting the B/C data.

In the rest of the paper, we are mostly sensitive to the high energy ($\gtrsim$ 10 GeV) description of the propagation. 
We check the reliability of our models, by performing a $\chi^2$-analysis and comparing the predicted B/C ratio with data from the most accurate surveys, namely, CREAM~\cite{Ahn:2008my}, ATIC~\cite{Panov:2007fe}, HEAO3~\cite{Engelmann:1990}, and CRN~\cite{Swordy:1990}, at $E\geq 3$ GeV. Results are reported in the last column of Table 1.

{\bf Energy loss:}
All the energy losses are computed within the Galprop code as described in Ref.~\cite{Strong:1998pw}.

Updated calculations of the ISRF~\cite{Porter:2006tb} have estimated a quite different emission in the inner region of the Galaxy, depending on the assumptions on the metallicity gradient. The picture in the outer region is, however, basically unchanged, and our results can be only very mildly affected.

The large-scale structure properties of the magnetic field are not extremely important as the turbulence properties in determining the diffusion. The strength is, on the other hand, crucial for the estimates of both synchrotron radiative emission and energy loss.
It turns out that the latter is a subdominant component of the energy loss term of Eq.~\ref{Eq:transport} in most regions of the Galaxy.
A precise estimate of the magnetic field strength is thus marginally relevant for X- and $\gamma$-ray emissions, while it becomes obviously very important when discussing radio and infrared signals.

The magnetic field strength can be estimated from pulsar data as~\cite{Han:2006ci}: 
$B=B_0 exp(-\frac{R-R_s}{R_B})$, with $R_s$ being the Sun-Galactic Center (GC) distance, $B_0=2.1\pm 0.3 \,\mu$G and $R_B=8.5 \pm 4.7$ kpc (similar results from extragalactic sources~\cite{Brown:2007qv}).
Note that polarization observations refer only to the line-of-sight component of the magnetic field. 
Radio synchrotron measurements suggest higher values for the strength of the total field $B$, namely, 6 $\mu$G near the Sun and about 10 $\mu$G in the inner Galaxy (outside the GC), assuming equipartition between the energy densities of magnetic fields and cosmic rays~\cite{Beck:2008ty} (this result is fairly in agreement with observations through the Zeeman splitting of atomic and molecular lines~\cite{Heiles:2005hr}). The radial scale length of the equipartition field is of about 12 kpc.
On the other hand, analysis of the WMAP synchrotron foreground data (plus some assumptions on the CR distribution and turbulence model) can lead to~\cite{MivilleDeschenes:2008hn} $B_0 = 3 \mu$G, $R_B$ = 11 kpc, and $B_{turb} /B_0 = 0.57$, not far from the estimate though rotation measures of pulsars.

We consider the benchmark case $B=5\cdot \exp[-(R-R_s)/10\,{\rm kpc}-|z|/2\,{\rm kpc}]\,\mu$G, with $R_s=8.5$ kpc.

\section{Source terms}
\label{sec:source}

\subsection{Standard primary cosmic ray components}
\label{sec:CR}

There are strong indications that the main mechanism of acceleration for primary Galactic CRs,
up to energies of 100~TeV or so, is the scattering of CR particles with the strong shock wave fronts produced by supernova remnants (SNRs) in the circumstellar medium~\cite{Aharonian:2006ws}. 
We will assume the primary CR source to be in the form:
\begin{equation}
   Q_i^p(R,z,E) \propto R^{\alpha_s} \exp\left(-\frac{R}{R_s} \right) 
                                                                \exp\left( -\frac{|z|}{z_s}\right) E^{-\beta_{inj,i}} \,,
\label{Eq:spectra}
\end{equation}
where $\alpha_s \simeq 2.35$, the radial length scale  $R_s \simeq 1.528$~kpc, and the 
vertical cutoff $z_s=0.2$~kpc confines the source distribution to the Galactic plane.
Neglecting discreteness and time variation effects, which could be eventually considered in 
connection to young nearby SNRs, the spatial part  of the source function follows the mean 
SNR distribution in the Galaxy as derived from radio pulsar population 
surveys~\cite{Lorimer:2003qc}.
This functional form, which is also in rough agreement with the gas distribution, traces the
Galactic Type~II supernova distribution and is highly suppressed in the Galactic bulge region. 
A contribution to primary cosmic rays from Galactic Type Ia supernova could be potentially 
relevant in the inner Galaxy (as one finds extrapolating from the distribution of old disc 
stars~\cite{Ferriere:2001rg}), while it's certainly very subdominant in the local environment;
since we will be mainly concerned with local observables, we are not sensitive to this
this contribution and simply disregard it.

Regarding the energy dependence in the primary CR source function, the theory of first order 
Fermi acceleration at astrophysical shocks predicts, for relativistic particles, a power-law 
spectral behavior, with the injection spectral index $\beta_{inj} \simeq 2$ in the limit of strong 
shocks~\cite{Blandford:1987pw}.  Assuming that the local measurements are representative of
the primary CR densities at the Sun galactocentric radius, we deduce the injection indices as 
well as source function normalizations matching the primary spectra after propagation to
the local measurements. For cosmic-ray nuclei, at high energy, spatial diffusion is the dominant
term in the propagation equation Eq.~(\ref{Eq:transport}); the injection index can then be derived
simply subtracting the scaling of the spatial diffusion coefficient $\alpha$ from the spectral index
of the locally measured flux, $\beta_{nuc}=2.7$, namely $\beta_{inj,nuc}=\beta_{nuc}-\alpha$, with 
$\alpha=1/3$ for a Komogorov diffusion, or $\alpha=1/2$ for a Kraichnan spectrum.
On the other hand, energy losses leads typically to a softer spectrum for high energetic primary 
electrons~\cite{Kobayashi:2003kp}. 
The most precise measurement of the $e^++e^-$ spectrum at high energy has been recently reported by the FERMI collaboration~\cite{Abdo:2009zk}.
The locally observed scaling $E^{-\beta_{e}}$, with 
$\beta_{e}\simeq 3.05$, can be reproduced assuming $\beta_{inj,e}\simeq 2.35$.
At enrgy above 1 TeV, HESS data~\cite{Collaboration:2008aaa} suggest the presence of an exponential cutoff. We multiply Eq.~\ref{Eq:spectra} by $\exp(-(E/5 {\rm TeV})^2)$, which is found to reproduce well the experimental data.
Finally, for both primary nuclei and electrons, we allow for a low energy break in the spectral 
indices, as suggested by the diffuse synchrotron and soft $\gamma$-ray emissions of the 
Galaxy~\cite{Strong:1998fr}. 

\begin{figure}[t]
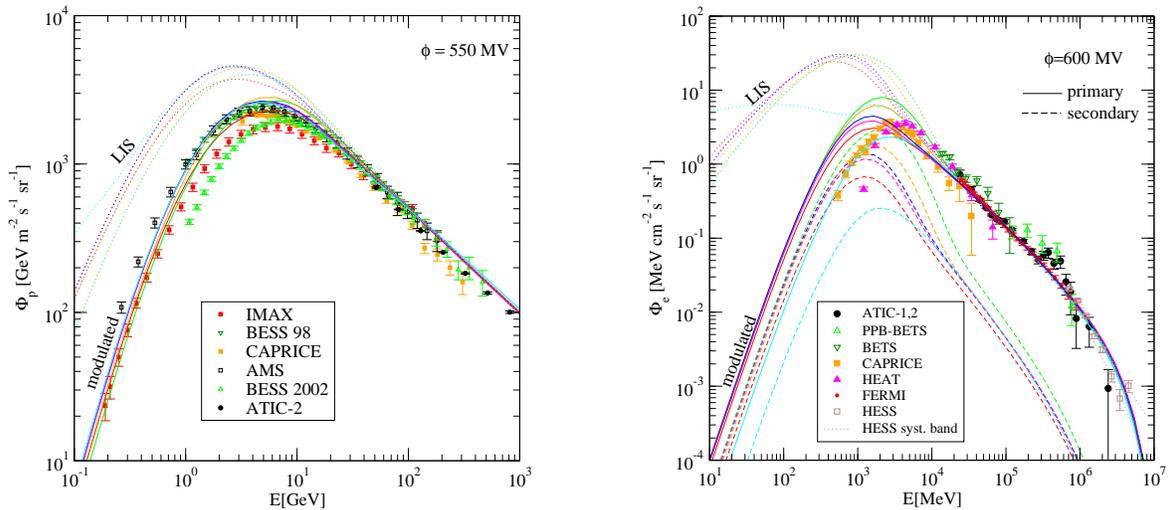

 \begin{minipage}[htb]{7cm}
   \centering
   \includegraphics[width=7.cm]{fig2a.eps}
 \end{minipage}
 \ \hspace{10mm} \
 \begin{minipage}[htb]{7cm}
   \centering
   \includegraphics[width=7.cm]{fig2b.eps}
 \end{minipage}
    \caption{Local proton ({\it Left Panel}) and electron ({\it Right Panel}) spectra. Data are compared to the benchmark models of propagation B0 (blue), B1 (green), B2 (red), B3 (orange), B4 (cyan), and B5 (magenta), described in Table 1.}
\label{fig:CRspectra}
 \end{figure}

The local fluxes, $\Phi_i= N_i\cdot E_i^2\,v_i/(4\pi) $, for protons and primary and secondary electrons as computed for the six 
benchmark models selected in the previous Section are shown in Fig.~\ref{fig:CRspectra}.
The corresponding injection spectral indices are reported in Table 1.

Several authors have questioned the assumption that local measurements are
representative of the CR density in the Galaxy, see e.g.~\cite{Pohl:1998ug}. Substantially
different spectral indices at high energy or normalizations have been sometimes
invoked for electron and/or proton spectra, e.g. to address~\cite{Strong:2004de} the excess 
above 1~GeV reported by EGRET in the observations of the diffuse gamma-ray emission from the 
Galaxy~\cite{Hunger:1997we}. This scenario seems now not strongly supported by preliminary
data  from the FERMI gamma-ray telescope. 
The spectrum of electrons with energy $E\gtrsim 1$ TeV is highly affected by the local picture, since they travel short distances due to the energy losses. Recent HESS data~\cite{Collaboration:2008aaa} suggest the existence of at least one source of CR electrons within $\sim 1$~kpc. On the other hand, 
the steepening in the energy spectrum without a clear hardening before seems to exclude strong contributions from local sources.
A different spectrum for protons is even more unlikely, since the scale over which protons diffuse
is much larger than for high energy electrons, and their density is not sensibly affected by 
a spatially inhomogeneous energy loss term. We will not consider further this possibility; all results
below are derived under the hypothesis that primary CR components can be normalized to 
the locally measured fluxes.

\subsection{Additional astrophysical sources for electron/positron cosmic ray components}
\label{sec:CR2}

Secondary electrons and positrons derive from the decays of charged pions produced in the
interaction of primary cosmic rays with the ISM along their propagation in the
Galaxy. Their spectral index at the source is equal to the spectral index of primaries after 
propagation , i.e. close to $\beta_{nuc} = 2.7$, which is larger than the injection spectral 
index for electrons,  $\beta_{inj,e}\simeq 2.35$. The ratio of secondary to primary electrons is thus expected to decrease as the energy increases. To reverse this trend, and fit the sharp raise in 
the positron fraction detected by PAMELA~\cite{Adriani:2008zr} above 10~GeV and up to 
100~GeV, it seems unavoidable to introduce an extra electron/positron source with harder 
spectrum. Several possibilities have been discussed. Pulsars are well-motivated astrophysical candidates, see, e.g.,~\cite{Aharonian,Hooper:2008kg,Profumo:2008ms}. We consider here
the alternative proposal of a possible production of secondary $e^+/e^-$ from hadronic interaction
within CR sources. Being injected in the same region where cosmic rays
are accelerated, they would have a harder spectrum than secondaries in the ISM, leading to a secondary to primary ratio increasing with energy~\cite{Blasi:2009hv}. 

We model this additional component assuming that the spectrum at the sources is described by a power-law plus an exponential cutoff: $E^{-\beta_{inj,sas}}\cdot \exp(-E/E_c)$. 
We consider a Bohm diffusion inside the source, which implies a spectral index $\beta_{inj,sas}=\beta_{inj,nuc}-1$ (at high energy). The spatial part of this extra source is assumed to be the same as for 
standard primary components. The normalization follows instead from the requirement that 
PAMELA data can be fitted when including this additional term. 

Although the physical insight for this picture is different from the case in which the enhancement 
in the positron fraction is due to one or few nearby pulsars, from the point of view of testing the 
scenario through radiative emission the two cases are hardly distinguishable. In both cases local 
sources dominate the signal, and in both cases these sources are confined in the thin vertical
layer where standard primary sources are confined. The discussion we present below for 
secondaries at the sources is then readily extendable to the pulsar scenario.

\subsection{Component from dark matter annihilations or decays}

A further possibility is that the extra component needed to explain the rise in the positron fraction
is an exotic term due to dark matter in the Galactic halo. 
There are two possibilities: WIMP dark matter particles are stable but can annihilate in pairs 
injecting a given species $i$;  the source term associated to this process is given by:
\be
Q_i^a(r,E)=(\sigma_a v)\,\frac{\rho(r)^2}{2\,M_{\chi}^2} \times \frac{dN_i^a}{dE}(E) \;,
\label{eqQ}
\ee
where $\rho(r)$ is the Milky Way halo mass density profile, assumed for simplicity to depend 
only on the spherical coordinate $r$, $M_{\chi}$ the mass of the dark matter particle, 
$\sigma v$ the pair annihilation rate for typical velocities of dark matter particles in the Galactic
halo (namely, in the zero temperature limit, as opposed to the finite temperature regime which 
applies in the early Universe), and $dN_i^a/dE(E)$ is the number of particles $i$ emitted 
per annihilation in the energy interval $(E,E+dE)$. The second possibility is that dark matter 
particles have a long but finite lifetime, and the species $i$ is injected in dark matter decays (for the interpretation of the PAMELA anomaly in terms of decaying DM, see, e.g.,~\cite{Yin:2008bs,Chen:2008qs,Ibarra:2008jk,Nardi:2008ix});
in this case the source function takes the form:
\be
Q_i^d(r,E)=\Gamma_d\,\frac{\rho(r)}{M_{\chi}} \times \frac{dN_i^d}{dE}(E) \;,
\label{eqQ2}
\ee
where $\Gamma_d$ is the decay rate and $dN_i^d/dE(E)$ is the number of particles $i$ 
emitted per decay in $(E,E+dE)$.

The distribution of dark matter in the Galaxy is rather poorly known, and one
has to rely on large extrapolations. One possibility is to take N-body simulations of hierarchical 
clustering in cold dark matter cosmologies as a guideline. Numerical results indicate that 
dark matter halos can be described by density profiles that are sharply enhanced towards the 
Galactic center;  there is still an on-going debate regarding how cuspy the profiles are, while,
from the observational point of view,  a tension has been often claimed between the profiles 
found in the numerical simulations and the dark matter density distribution as inferred from
circular velocity or velocity dispersion maps, especially in small or low brightness objects.
For what concerns our analysis, introducing a dark matter model with large density towards
the center of the Milky Way would trigger discussion on signals generated in the central portion 
of the Galaxy. In line with other recent analysis, e.g., Refs.~\cite{Cholis:2008wq,Zhang:2008tb,Barger:2009yt,Bertone:2008xr,Cirelli:2009vg}, strong constraints would arise. However these limits
are strongly model dependents, relying on several extrapolations with respect to quantities
that are measured and or tested with local observables, first of all the one on the dark matter
density profile itself. We take here a different route and try to sketch results that depend as weakly
as possible on extrapolations of local quantities. For the dark matter profile we assume a 
functional form with a large core radius in the central region of the Galaxy, the Burkert 
profile~\cite{Burkert:1995yz}:
\be
\rho(r)=\frac{\rho_0}{(1+r/r_c)\,(1+(r/r_c)^2)} \,\,\,\,\,,
\label{Eq:Burkert}
\ee
where $r$ is the distance from the GC; profile normalization $\rho_0=0.84$~GeV~cm$^{-3}$ and 
the core radius $r_c=11.7$~kpc, corresponding to a local halo density of 
$\rho(r_0)= 0.34$~GeV~cm$^{-3}$, are obtained for a model of the Milky Way halo with virial mass
and concentration parameter being, respectively, $M_{vir} = 1.3 \times 10^{12}\msun$ 
and $c_{vir} = 16$. This model matches a whole set of available dynamical 
informations~\cite{Edsjo:2004pf}, 
including, among others, constraints from the motion of stars in the Sun's neighbourhood, total 
mass estimates following from the motion of the outer satellites, and the Milky Way rotation curve.
The Burkert profile, which was originally introduced as a phenomenologically model from 
dynamical observations, from the theoretical point of view has been discussed as the equilibrium
configurations from cold dark matter models in case the baryon infall happens with a sensible 
exchanged of angular momentum between baryons and dark matter~\cite{elzant}.

Note that for the dark matter density profile in Eq.~(\ref{Eq:Burkert}) the gradient of the density
profile in the region where most of the signal we will consider originates is very modest. 
Hence, in practice, the spatial signature of the annihilation source function, due to the scaling 
of the source function Eq.~(\ref{eqQ}) with the number density of WIMP pairs, i.e. with 
$\rho^2(r)$, is hardly distinguishable from that of the decay source function, scaling simply
with the number density of dark matter particles, i.e. with $\rho(r)$.

Unresolved Galactic subhalos could, in principle, sligthly modify our conclusions, depending on their description.
In particular, if the spatial distribution of subhalos is antibiased with respect to the host halo mass distribution (as found, e.g., in the Via Lactea simulation~\cite{Kuhlen:2008aw}), the emission at mid-high latitudes would be enhanced and the conclusions would be strengthened (while they would be weakened in the opposite picture). On the other hand, the estimates of mass function, spatial distribution and concentration of subhalos have great uncertainties and, taking a conservative approach, we disregard the contribution of substructures to the total DM density.

If the dark matter particle mass is not known, 
it will be difficult to discriminate the two cases even from the spectral shape of the sources. To 
reproduce the PAMELA positron rise, a positron dark matter source with hard spectrum is required;
this is possible in case of prompt annihilation or decays into leptons or weak gauge bosons. 
The second possibility is strongly constrained by available measurements of the antiproton cosmic 
ray flux (see, e.g.,~\cite{Cirelli:2008pk}), since $W^\pm$ and $Z^0$ are copious sources of 
antiprotons. We will consider annihilations in pairs into monochromatic $e^+/e^-$ 
($E_e \simeq M_{\chi}$) and monochromatic $e^+/e^-$ from a two-body final state in the
decay ($E_e \simeq M_{\chi}/2$), as well cascades initiated by annihilations or decays into $\mu^+/\mu^-$ and
$\tau^+/\tau^-$. 
In the cases with $e^+/e^-$ and $\mu^+/\mu^-$ as final states of annihilation/decay, the resulting $e^+/e^-$ source comes together with a gamma-ray component
stemming from radiative processes at emission; we consider the model independent part
of the final state radiation (FSR) terms (limit of $M_{\chi} \gg m_e$) and neglect eventual model 
depend terms for specific WIMP pair annihilation cases (sometimes referred to as virtual internal 
bremsstrahlung~\cite{Bergstrom:2004cy,Birkedal:2005ep}) or for  the case of a DM 
candidate annihilating into new light particles which in turn generate $e^+/e^-$, 
e.g.,~\cite{Bergstrom:2008ag} (in this last case the FSR component may be reduced with 
respect to what we are implementing in the rest of the paper).
For the case mediated by $\tau^+/\tau^-$, on top of the FRS term, 
there is a substantial $\gamma$-ray flux induced by neutral pion production and their subsequent 
decay into 2~photons; this term is accounted for linking to the Pythia Montecarlo simulations as 
provided by the \ds\ package.

The $\gamma$-ray signal is directly related to the injection source of Eqs.~\ref{eqQ} and \ref{eqQ2}.
The computation of the radiative emissions requires the solution of Eq.~\ref{Eq:transport} with the $e^+/e^-$ injection source described in Eqs.~\ref{eqQ} and \ref{eqQ2}. We perform the calculation of 
this term with the Galprop package, analogously to the previous contributions.  
Relevant formulas to infer some scalings of the results (in case propagation is treated in an 
approximate form) are described, e.g., in Sec.~3 of Ref.~\cite{Regis:2008ij}.

\section{Signal vs Background}
\label{sec:signal}
\subsection{Electron/Positron spectrum}

\begin{figure}[t]
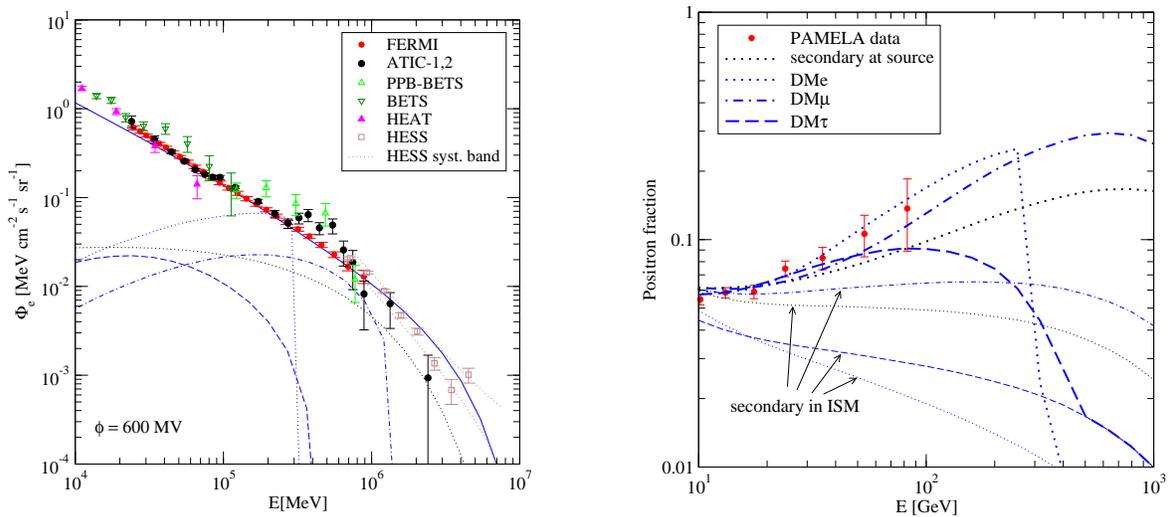

 \begin{minipage}[htb]{7cm}
   \centering
   \includegraphics[width=7.cm]{fig3a.eps}
 \end{minipage}
 \ \hspace{10mm} \
 \begin{minipage}[htb]{7cm}
   \centering
   \includegraphics[width=7.cm]{fig3b.eps}
 \end{minipage}
    \caption{{\it Left Panel}: Local electron+positron spectrum in the "conventional" model B0. We show primary CR electrons (blue solid), secondary CR electrons + positrons produced at the source (black dotted), and electrons + positrons induced by WIMP annihilation in the model DM$e$ (blue dotted), DM$\tau$ (blue dashed), and DM$\mu$ (blue dashed-dotted). {\it Right Panel}: Local positron fraction above 10 GeV in the "conventional" model B0. Best-fit curves including the contribution of the positron sources shown in the left panel are plotted with thick lines. They are obtained by allowing mild variations in the normalization and spectral index of primary and secondary electrons/positrons spectra, as described in the text. Contribution from secondary positrons produced in the ISM is shown with thin lines. Line-styles as in the left panel.}
\label{fig:LISepem}
 \end{figure} 

For sake of clearness, we concentrate our analysis on three benchmark WIMP scenarios, named 
DM$e$, DM$\tau$, and DM$\mu$, and summarized in Table 2. Injecting monochromatic $e^+/e^-$ yields, the model DM$e$ induces a very steep electron/positron spectrum. This is true also for the model DM$\mu$, while for the model DM$\tau$, which assumes $\tau^+/\tau^-$ are produced, the spectrum is quite hard but
significantly smoother. A dark matter scenario fitting the PAMELA positron excess typically lies in between these three cases.
Note that in the case of DM annihilations/decays producing such final states after a number of steps mediated by new exotic light particles, the final spectrum of $e^+/e^-$ is softened with respect to the picture considered in the following (to an extent depending on the number of steps and on the properties of the new particles).

{\small
\begin{table}[b]
\begin{center}
\begin{tabular}{|c|c|c|c|c|c|}
\hline
$\,\,\,\,\,\,\,\,\,\,\,$&$M_\chi$&$\sigma_a v$&annihilation&spatial&line\\
$\,\,\,\,\,\,\,\,\,\,\,$&[GeV]&$[{\rm cm^3 s^{-1}}]$& modes & profile&coding
\tabularnewline
\hline
\hline
DM$e$&300&$2.5\cdot 10^{-24}$&$e^+ e^-$&Burkert&dotted
\tabularnewline
\hline
\hline
DM$\tau$&400&$6.6\cdot10^{-24}$&$\tau^+ \tau^-$&Burkert&dashed
\tabularnewline
\hline
\hline
DM$\mu$&1500&$2.5\cdot10^{-23}$&$\mu^+ \mu^-$&Burkert&dashed-dotted
\tabularnewline
\hline
\end{tabular}
\end{center}
\caption{WIMP benchmark models. The annihilation rates are listed in the case of the "conventional" propagation model B0. }
\label{tabDM}
\end{table}

}

\begin{figure}[t]
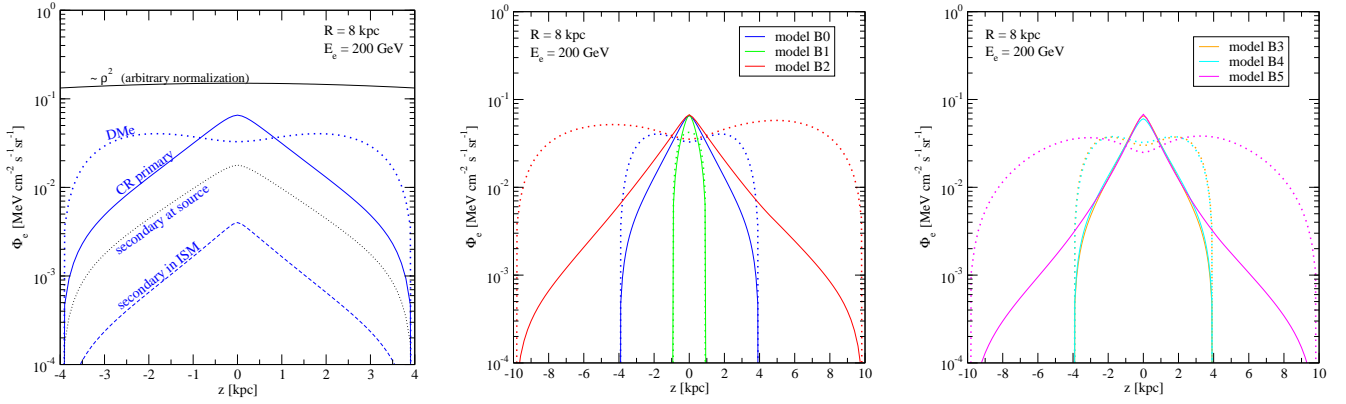

 \begin{minipage}[htb]{5.5cm}
   \centering
   \includegraphics[width=5.5cm]{fig4a.eps}
 \end{minipage}
 \ \hspace{1mm} \
 \begin{minipage}[htb]{5.5cm}
   \centering
   \includegraphics[width=5.5cm]{fig4b.eps}
 \end{minipage}
 \ \hspace{1mm} \
 \begin{minipage}[htb]{5.5cm}
   \centering
   \includegraphics[width=5.5cm]{fig4c.eps}
 \end{minipage}
    \caption{Electron vertical profile at $R=8$ kpc and $E=200$ GeV. {\it Left Panel}: Propagation model B0: We show primary CR electrons (blue solid), secondary CR $e^+ + e^-$ produced in the ISM (blue short dashed), secondary CR $e^+ + e^-$ produced at the source (black dotted), and $e^+ + e^-$ induced by DM annihilation in the model DM$e$ (blue thick dotted). For comparison, we plot the distribution of a source scaling as $\rho_{DM}^2$ (black solid), with an arbitrary normalization. {\it Central Panel}: The same of the left panel, but adding the propagation models B1 (green) and B2 (red) and considering only the contributions from primary electrons and DM induced electrons + positrons.
{\it Right Panel}: The same of the central panel, but in the propagation models B3 (orange), B4 (cyan), and B5 (magenta).}
\label{fig:vertprof}
 \end{figure} 

We start with the case of pair annihilating WIMPs.
The explanation of the PAMELA data requires a WIMP mass 
$M_{\chi}\geq 80$~GeV. WIMPs with $M_{\chi}< 1$~TeV and inducing a very hard spectrum of $e^+ / e^-$ are disfavoured by the FERMI data on $e^+ + e^-$ (assuming a power-law for the injection spectra of primary electrons). We consider the scenario of a WIMP annihilating directly into $e^+ / e^-$ as a toy benchmark case to show that for any DM model inducing a hard spectra of $e^+ / e^-$ fitting the PAMELA positron fraction, the associated diffuse emission is at a detectable level, independently on the mass (for $M_{\chi}\gtrsim 150$ GeV). 
A heavy WIMP (i.e., $M_{\chi}\gtrsim$ few hundreds of GeV) with the same final state would actually induce 
an IC emission which is already strongly constrained by EGRET data~\cite{Cirelli:2009vg}. For the DMe benchmark scenario, we consider a WIMP mass of 300~GeV. 
A light(i.e., $M_{\chi}\lesssim 300$ GeV) WIMP candidate 
annihilating into $\tau^+ \tau^-$ does not provide an explanation to the PAMELA data.
The $\gamma$-ray emission induced by a heavy (i.e., $M_{\chi}\gtrsim 500$ GeV) DM candidate with the same annihilation modes (and fitting the PAMELA excess) could violate constraints by ACT measurements of dwarf satellites~\cite{Essig:2009jx} and of the Galactic Ridge~\cite{Bertone:2008xr}. We refer to a sample 
case standing in middle of this mass interval, namely a WIMP mass of 400~GeV (DM$\tau$).
A WIMP annihilating into $\mu^+ \mu^-$ with mass $M_{\chi}> 1$~TeV can give a very good fit to, simultaneously, the PAMELA and FERMI data. In the benchmark case DM$\mu$, we consider $M_{\chi}= 1.5$ TeV.

We tune the annihilation cross section in order to fit the positron fraction measured by PAMELA above 10 GeV and the $e^+ + e^-$ local spectrum measured by FERMI. 
The fit is performed adding the dark matter component to 
the background contribution from standard primary CRs and the secondaries produced along
propagation, computed in the same propagation model configuration. We allow the normalization and spectral index of both primary and secondary CR electrons to vary with respect to the cases reported in Table 1 (within the ranges $\Delta \beta_{inj,e}=\pm 0.5$ and $\Delta \beta_{inj,nuc}=\pm 0.1$), reflecting some uncertainties on their modelling.
Within these assumptions, the models describing the DM component with the benchmark scenario DM$e$ cannot reproduce simultaneously the two sets of data. They would require some exotic assumptions for the primary injection spectra. Therefore, in this case we restrict the analysis to the fit of the PAMELA dataset only, requiring the $e^+ + e^-$ spectrum induced by DM$e$ alone to be consistent with the FERMI observations.

The best fit values for the DM$e$ scenario are: 
$\sigma_a v =(2.5,2.3,2.7,2.5,2.7,2.1)\cdot 10^{-24}\,{\rm cm}^3{\rm s}^{-1}$, for DM$\tau$ scenario are: $\sigma_a v =(6.6,4.2,6.1,8.0,6.6,5.4)\cdot 10^{-24}\,{\rm cm}^3{\rm s}^{-1}$, and for DM$\mu$ scenario are: $\sigma_a v =(2.5,2.0,2.3,2.8,2.8,2.4)\cdot 10^{-23}\,{\rm cm}^3{\rm s}^{-1}$  in the (B0, B1, B2, B3, B4, B5) propagation model, respectively.
For the DM$\tau$ and DM$\mu$ scenarios, all the models provide a good fit to the data.
The theoretical curves do not reproduce the data in an extremely precise way only for what concerns the positron fraction at energies below 20 GeV in the propagation models B1 and B3. In such cases, the production of secondary positrons from primary CR is enhanced 
(see Fig.~\ref{fig:CRspectra}b) and the background alone tends to be in conflict with PAMELA data at low energies.
We plot in Fig.~\ref{fig:LISepem} the obtained $e^+ + e^-$ spectrum and positron fraction of the three benchmark DM scenarios in the propagation model B0.

Note that, as expected, the best-fit values for the annihilation rates are significantly larger than the thermally averaged annihilation rate at the time of decoupling of WIMPs, namely, $\sigma_a v =3\cdot 10^{-26}\,{\rm cm}^3{\rm s}^{-1}$.
Many explanations have been proposed to motivate such mismatch, like, e.g., non thermal production mechanism, non-standard cosmology at the time of DM decoupling, and Sommerfeld enhancement. 

As we already mentioned, in the case of Burkert DM profile, the differences in the spatial distribution of a signal scaling with $\rho^2$, as in the WIMP case, or with $\rho$, as in the decaying DM case, are very mild. 
Therefore, our results can be rephrased in term of benchmark decaying DM scenarios with decaying modes analogous to the annihilation modes introduced above, and with mass scale doubled. The corresponding decay rate can be approximately estimated by $\tau \simeq (\sigma_a v)^{-1}\cdot M_{\chi}/\rho_0 \simeq 3.3 \cdot 10^{26} (M_{\chi}/100 {\rm GeV})\cdot (10^{-24}\,{\rm cm}^3{\rm s}^{-1}/(\sigma_a v))$, where $M_{\chi}$ is the mass of the WIMP in the annihilating DM scenario.

As discussed in the previous Section, for comparison, we consider also the case of an extra non-standard component due to the production of secondary $e^+/e^-$ inside CR sources. In Fig.~\ref{fig:LISepem}, we plot the $e^+ + e^-$ spectrum and positron fraction in the propagation model B0 for a population injected with $\beta_{inj,sas}=1.36$, an energy-cutoff at $E_c=1$~TeV and normalization tuned to the best fit value for the PAMELA and FERMI data (again allowing mild variations in the normalization and spectral index of primary and secondary electrons).

The absence of sharp features in the $e^+ + e^-$ spectrum detected by FERMI implies that the contribution of the source of positrons responsible for the PAMELA excess to the total $e^+ + e^-$ local flux or to the associated diffuse emission, can be hardly singled out.  
If the spatial distribution of the sources of such "exotic" components traces the CR sources, the last sentence would be true everywhere in the Galaxy.

In Fig.~\ref{fig:vertprof}, we plot the vertical profiles of the electron number density distributions, at 
the local radial distance  $R=8$~kpc and $E=200$~GeV. For this slice of the Galaxy, at rather large
distance from the GC, the determination of propagation model parameters as derived by matching 
the LIS of nuclei is rather robust. We are focusing on some typical energy at which
electron and positron sources relevant for the raise in the positron fraction are also a significant contribution to the total population of $e^+ + e^-$.
In Fig.~\ref{fig:vertprof}a we consider the propagation model B0, and plot the vertical profile of CR primary electrons (solid), secondary $e^+ + e^-$ produced in the ISM (short-dashed), secondary $e^+ + e^-$ injected at the source (dashed-dotted), and $e^+ + e^-$ flux induced by WIMP annihilations in the DM$e$ scenario (dotted).
All these cases but the latter follow a distribution which is mainly confined to the disc (although broadened by the diffusion). The DM-induced component is instead much flatter (we plot for comparison the profile of the DM injection source $\propto \rho_{DM}^2$). 
It is the dominant component at intermediate and large z. We thus expect the associated radiative emission to dominate at intermediate and high latitudes.

In order to understand how this conclusion is dependent on the propagation model considered, we show the cases of the propagation model B1 and B2 (plus again B0, for comparison) in Fig.~\ref{fig:vertprof}b, and of B3, B4, and B5 in Fig~\ref{fig:vertprof}c. In these figures, we do not plot the vertical profiles of secondary $e^+ + e^-$. Their shapes are analogous to the CR primary electrons profile and the rescaling factor is roughly the same as in Fig.~\ref{fig:vertprof}a.
Note from Fig.~\ref{fig:vertprof}b that, as expected, as the boundary of propagation $z_h$ increases (decreases), the region at which the DM-induced component is dominant becomes larger (smaller).  
From Fig.~\ref{fig:vertprof}c, we conclude that at high energies the effect of convection (model B3) on the shape of the vertical profile is negligible. The same conclusion applies also to the effect of varying the spectral index of the diffusion coefficient (model B4).
In the model B5, the population of electrons induced by DM at high $z$ is mildly reduced with respect to the model B2 since the spatial diffusion coefficient is increasing with $z$, and electrons and positrons 
are less efficiently confined.  

\begin{figure}[t]
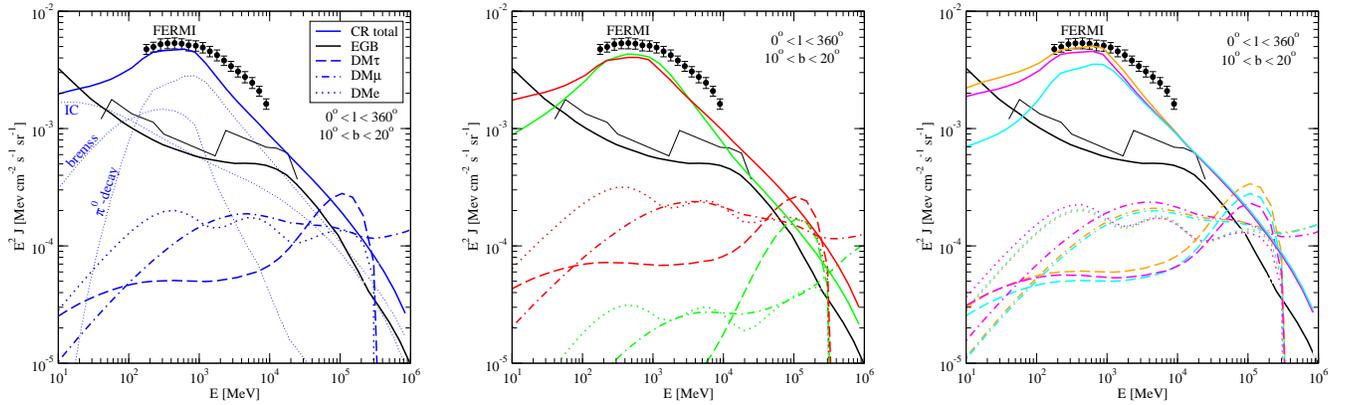

 \begin{minipage}[htb]{5.5cm}
   \centering
   \includegraphics[width=5.5cm]{fig5a.eps}
 \end{minipage}
 \ \hspace{1mm} \
 \begin{minipage}[htb]{5.5cm}
   \centering
   \includegraphics[width=5.5cm]{fig5b.eps}
 \end{minipage}
 \ \hspace{1mm} \
 \begin{minipage}[htb]{5.5cm}
   \centering
   \includegraphics[width=5.5cm]{fig5c.eps}
 \end{minipage}
    \caption{$\gamma$-ray diffuse spectrum at intermediate latitudes ($10^{\circ}<b<20^{\circ}$), integrated over longitudes $0^{\circ}<l<360^{\circ}$ and compared to the FERMI preliminary data~\cite{FERMI:preliminary}. {\it Left Panel}: Emission in the propagation model B0. The CR (primary+secondary) spectra associated to $\pi^0$-decay, IC, and bremsstrahlung are shown by thin dotted lines. The thick solid blue line is the sum of the three components. The solid black line shows the extragalactic background in the model described in the text (thick) and fitted from EGRET data~\cite{Strong:2004de} (thin). The IC + FSR emission associated to the WIMPs DM$e$ and DM$\mu$ are shown by thick dotted and thick dashed-dotted lines, respectively. The IC + $\gamma$-ray from $\pi^0$-decay signals induced by the WIMP DM$\tau$ are shown by thick dashed lines.   {\it Central Panel}: Emission in the propagation models B1 (green) and B2 (red). Same line styles of the left panel. {\it Right Panel}: The same of central panel, but for the propagation models B3 (orange), B4 (cyan), and B5 (magenta).}
\label{fig:emissint}
 \end{figure}

\subsection{$\gamma$-ray emission}

The discussion in the previous Section pointed out that, in order to detect a DM-induced signal in the diffuse emission of the Galaxy, intermediate and high latitudes are the best targets.

\begin{figure}[t]
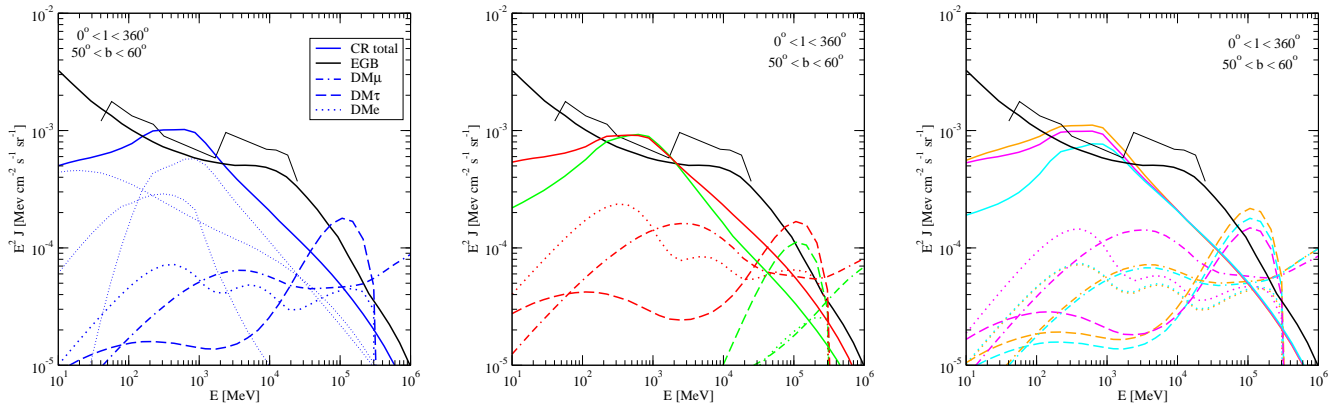

 \begin{minipage}[htb]{5.5cm}
   \centering
   \includegraphics[width=5.5cm]{fig7a.eps}
 \end{minipage}
 \ \hspace{1mm} \
 \begin{minipage}[htb]{5.5cm}
   \centering
   \includegraphics[width=5.5cm]{fig7b.eps}
 \end{minipage}
 \ \hspace{1mm} \
 \begin{minipage}[htb]{5.5cm}
   \centering
   \includegraphics[width=5.5cm]{fig7c.eps}
 \end{minipage}
    \caption{$\gamma$-ray diffuse spectrum at high latitudes ($50^{\circ}<b<60^{\circ}$) integrated over longitudes $0^{\circ}<l<360^{\circ}$. Line styles and colors as in Fig.~\ref{fig:emissint}.}
\label{fig:emisshigh}
 \end{figure}

At high latitudes, the diffuse extragalactic gamma-ray background (EGB) is expected to become 
the dominant background component.
To estimate the level of the extragalactic emission in the FERMI preliminary data~\cite{FERMI:preliminary} reported in Fig.~\ref{fig:emissint}, we rely on the EGRET data and we consider the fit obtained in Ref.~\cite{Strong:2004de} (upper black curve). The sharp increase in sensitivity to point sources of the FERMI telescope with respect 
to EGRET, may, on the other hand, lower significantly such term. In three months of observations, FERMI has already detected an amount of individually resolved active galactic nuclei (which are believed to be the main component of the EGB) corresponding to $\sim$ 7\% of the EGRET extragalactic diffuse gamma-ray background~\cite{Abdo:2009wu}.
We consider a model for the contribution of unresolved blazars as in Ref.~\cite{Ullio:2002pj} (lower black curve in Fig.~\ref{fig:emissint} and \ref{fig:emisshigh}), estimating the FERMI point source sensitivity as $1.6\cdot 10^{-9} {\rm cm^{-2} s^{-1}}$, roughly corresponding to 3 years of observations.
Another crucial ingredient to estimate the diffuse extragalactic radiation is absorption of gamma-rays 
at high energies, mainly due to pair production on the extragalactic background light emitted by 
galaxies in the ultraviolet, optical and infrared bands. We consider the parametrization of this effect
in Ref.~\cite{Primack:2000xp}, as derived in the context of the $\Lambda$CDM cosmological model.

In Fig.~\ref{fig:emissint}, we plot the $\gamma$-ray diffuse spectrum at $10^{\circ}<b<20^{\circ}$, integrated over longitude ($0^{\circ}<l<360^{\circ}$), and compared to the FERMI preliminary data.
These measurement do not confirm the EGRET excess in the GeV energy range, with the level of the detected diffuse flux being significantly reduced.
In Fig.~\ref{fig:emissint}a we show the case of the "conventional" propagation model B0.
In the previous Section, in order to determine the best fit values of the annihilation cross sections, we allowed the normalization and spectral index for primary and secondary CR electrons/positrons to vary with respect to the benchmark cases reported in Table 1 (see Fig.~\ref{fig:LISepem}b). On the other hand, such mild variations in the spectra induce negligible variations in the associated diffuse emissions.  Thus, in the following, for sake of clearness, we consider the background CR emission as given by the benchmark models of Table 1. 

The first remark is that the sum (blue solid line) of three CR components (blue thin dotted lines), namely, IC, bremsstrahlung, and $\pi^0$-decays, plus the extragalactic background contribution (black solid line), can approximately account for the measured flux at $E\leq 10$ GeV (note that propagation models
have not been tuned to do so, while we are just extrapolating from the LIS of nuclei).
Exotic components, claimed in order to explain the EGRET excess, are now significantly constrained, at least at mid-latitudes\footnote{Other observations reported by the FERMI LAT telescope (e.g., Vela pulsar~\cite{Abdo:2008ef}) go in the same direction, namely, reporting a reduced flux at GeV energies with respect to the EGRET observations. The current most likely interpretation of the EGRET excess is thus an instrumental bias. This would imply that a significant contribution from exotic components at 
few GeV is severely constrained in any portion of the sky.}.

In the same plot one can see that the $\gamma$-ray flux induced by our benchmark DM models 
is more than one order of magnitude smaller than the detected flux at $E\leq 10 $ GeV, while it becomes comparable to or higher than the background at $E\gtrsim$ 100 GeV. At such energies, both the IC and FSR signals are relevant in the model DM$e$ (thick dotted line) and DM$\mu$ (thick dashed-dotted line), while in the model DM$\tau$ (thick dashed line) the flux is driven by the $\pi^0$-decay emission. 
Note that the in the DM$e$ scenario the final $e^+/e^-$ spectrum is very sharp and thus the peaks of the IC emission on CMB and on infrared and optical starlight are clearly visible, while in the DM$\mu$ and DM$\tau$ cases they are smoothed.

The $\gamma$-ray diffuse spectrum at intermediate latitudes for the propagation models B1 and B2 is shown in Fig.~\ref{fig:emissint}b.
As expected, the IC signal associated to the WIMP DM$e$ and DM$\mu$ is enhanced (reduced) than the CR signal for the model B2 (B1) with respect to the "conventional" model, making the WIMP scenario more easily detectable (undetectable).
Moreover the CR secondary positron spectrum in the model B2 (B1) is reduced (enhanced), see Fig.~\ref{fig:CRspectra}b. In the fit of the PAMELA excess, this leads to an enhancement (decrease) in the normalization of the DM-induced positron spectrum, and thus in the annihilation cross section, implying a higher (smaller) signal associated to $\pi^0$-decay and FSR.

From Fig.~\ref{fig:emissint}c, one can see that the level of CR diffuse spectra at high energy in the model B3, B4, and B5 is almost identical to the "conventional" case. The DM-induced emissions are also analogous.

All the CR benchmark scenarios considered in the discussion induce emissions which are fair in agreement with the FERMI preliminary data. We find a slight mismatch only in the model B1 at high energy, caused 
by the lower level of IC and bremsstrahlung radiations as follows from the fact that we are introduced
a smaller region for $e^+/e^-$ propagation and thus a smaller region for radiative emission, and in the model B4 at low energy, due to the choice of spectral indices (see the $e^+/e^-$ spectrum at 
$\sim$~GeV energies in Fig.~\ref{fig:CRspectra}b).

As already emphasized a few times in the paper, the CR emission is almost confined on the disc (i.e., it is elongated in the $R$ direction and rapidly decreasing along $z$). Therefore, the diffuse emission in 
$(b,l)$ coordinates (namely, the flux integrated along the line of sight labeled by the angles $(b,l)$) tends to have a flatter longitudinal profile as the latitude increases.
This is true also for the DM components (being the DM profile and the ISRF distribution, relevant for IC emission, decreasing with $z$), but to a smaller extent, especially at intermediate latitudes.  
We expect therefore larger ratios for the DM signal to CR background at 
longitudes close to the center of the Galaxy; on the other hand, our procedure for estimating parameters in the propagation  model is driven by local data, and thus our extrapolations are more reliable outside the inner region.   

\begin{figure}[t]
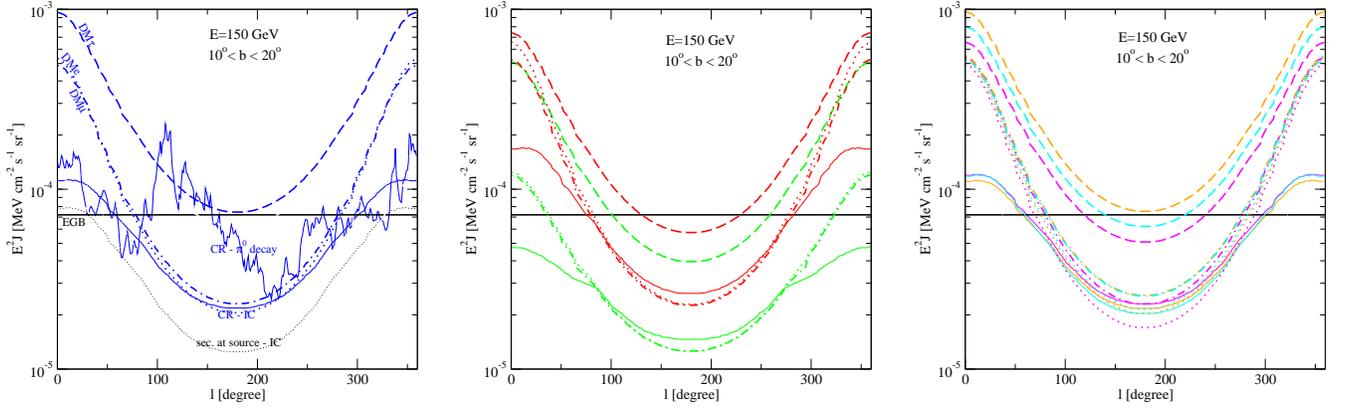

 \begin{minipage}[htb]{5.5cm}
   \centering
   \includegraphics[width=5.5cm]{fig6a.eps}
 \end{minipage}
 \ \hspace{1mm} \
 \begin{minipage}[htb]{5.5cm}
   \centering
   \includegraphics[width=5.5cm]{fig6b.eps}
 \end{minipage}
 \ \hspace{1mm} \
 \begin{minipage}[htb]{5.5cm}
   \centering
   \includegraphics[width=5.5cm]{fig6c.eps}
 \end{minipage}
    \caption{Longitudinal profile of the $\gamma$-ray diffuse emission at intermediate latitudes ($10^{\circ}<b<20^{\circ}$) and $E=150$ GeV. {\it Left Panel}: Emission in the propagation model B0. The CR profiles associated to $\pi^0$-decay and IC are shown by thin solid lines. The solid black line shows the extragalactic background in the model described in the text. The profile of the IC emission associated to secondary $e^+ + e^-$ produced at the source is shown by the black dotted line. The IC + FSR emissions associated to the WIMPs DM$e$ and DM$\mu$ are shown by thick dotted and dashed-dotted lines, respectively. The IC + $\gamma$-ray from $\pi^0$-decay signals induced by the WIMP DM$\tau$ are shown by thick dashed lines. {\it Central Panel}: Emission in the propagation models B1 (green) and B2 (red). Line styles as in the left panel. {\it Right Panel}: The same of central panel, but for the propagation models B3 (orange), B4 (cyan), and B5 (magenta).}
\label{fig:lonprofint}
 \end{figure}

In order to show that the detectability of a DM component inferred from Fig.~\ref{fig:emissint} is not only related to the picture in the central region of the Galaxy, in Fig.~\ref{fig:lonprofint} we plot the longitudinal profile for the diffuse emission at $E=150$ GeV and for $10^{\circ}<b<20^{\circ}$ (namely, the same latitudes of Fig.~\ref{fig:emissint}).  
Note that the diffuse component induced by DM dominates over a large range of longitudes for any benchmark scenarios. The picture is completely different with respect to the signal induced by secondary $e^+ + e^-$ produced at the source, which is subdominant everywhere, having a spatial profile of injection identical to the primary sources. 
The CR emission associated to $\pi^0$-decay is shown only in the "conventional" case (Fig.~\ref{fig:lonprofint}a). At high energy, it is analogous for all the models of propagation, being mostly generated close to the disc. It is highly irregular due to the variation of the gas density. The longitudinal shape of the $\pi^0$-decay signal in the southern hemisphere ($-20^{\circ}<b<-10^{\circ}$) would be different but with an overall size roughly analogous.
From the figures, one can see that the longitudinal shapes of the signals are analogous for all the benchmark propagation models.

\begin{figure}[t]
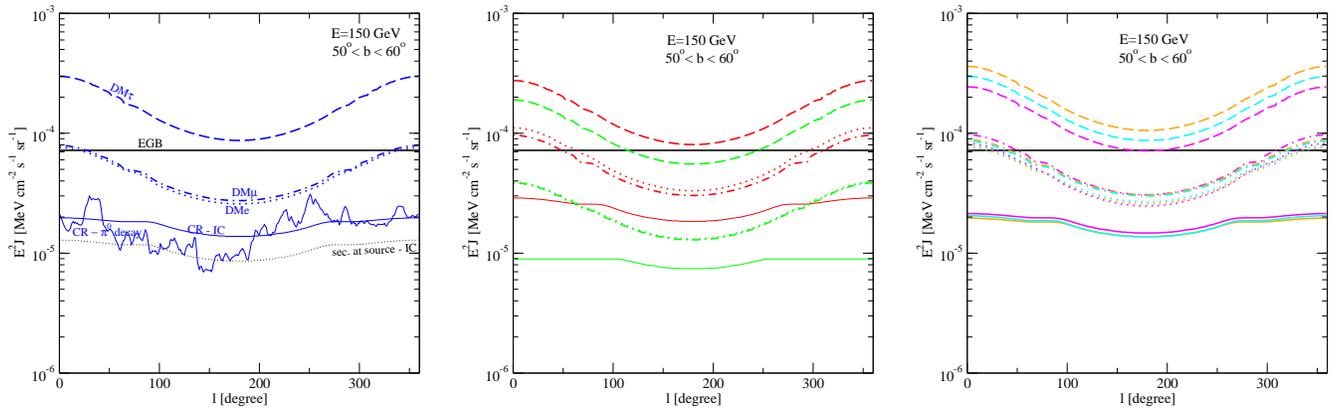

 \begin{minipage}[htb]{5.5cm}
   \centering
   \includegraphics[width=5.5cm]{fig8a.eps}
 \end{minipage}
 \ \hspace{1mm} \
 \begin{minipage}[htb]{5.5cm}
   \centering
   \includegraphics[width=5.5cm]{fig8b.eps}
 \end{minipage}
 \ \hspace{1mm} \
 \begin{minipage}[htb]{5.5cm}
   \centering
   \includegraphics[width=5.5cm]{fig8c.eps}
 \end{minipage}
    \caption{Longitudinal profile of the $\gamma$-ray diffuse emission at high latitudes ($50^{\circ}<b<60^{\circ}$) and $E=150$ GeV. Line styles and colors as in Fig.~\ref{fig:lonprofint}.}
\label{fig:lonprofhigh}
 \end{figure}

In Figs.~\ref{fig:emisshigh} and \ref{fig:lonprofhigh}, we repeat the same analysis at higher latitudes, namely, $50^o<b<60^o$.
As expected from the analysis of the previous Section, in this case we find a larger ratio between the signal induced by DM to the emission associated to CRs. On the other hand, the EGB starts to play
a major role.
The detection prospects for the WIMP scenario DM$\tau$ are again very favourable in all the propagation models. The emission induced by DM$e$ and DM$\mu$ is also detectable, being, roughly, of the same level of the sum of the backgrounds at $E\gtrsim 100$ GeV. This is no longer true at higher latitudes, where the EGB takes over and such emission becomes too faint to give a clear signature.
Fig.~\ref{fig:lonprofint}, shows that, as explained in the discussion above, the longitudinal profiles become flatter than at lower latitudes. The emissions come mostly from the local region and therefore these predictions can be assumed as rather robust.

Note that the enhancement in the DM-induced IC emission in the propagation models with $z_h=10$ kpc (B2 and B5) with respect to the "conventional" case ($z_h=4$ kpc) is more significant than at intermediate latitudes, and viceversa for the model B1.
The B2 case is more favourable than the B5 model; in the latter the $e^+/e^-$ population is 
slightly depleted at large $z$ since the spatial diffusion coefficient increases in such region.
The predictions in the models B3 and B4 are again analogous to the "conventional" case.

\begin{figure}[t]
 \ \hspace{-35mm} \
 \begin{minipage}[htb]{5.5cm}
   \centering
   \includegraphics[width=5.5cm,angle=90]{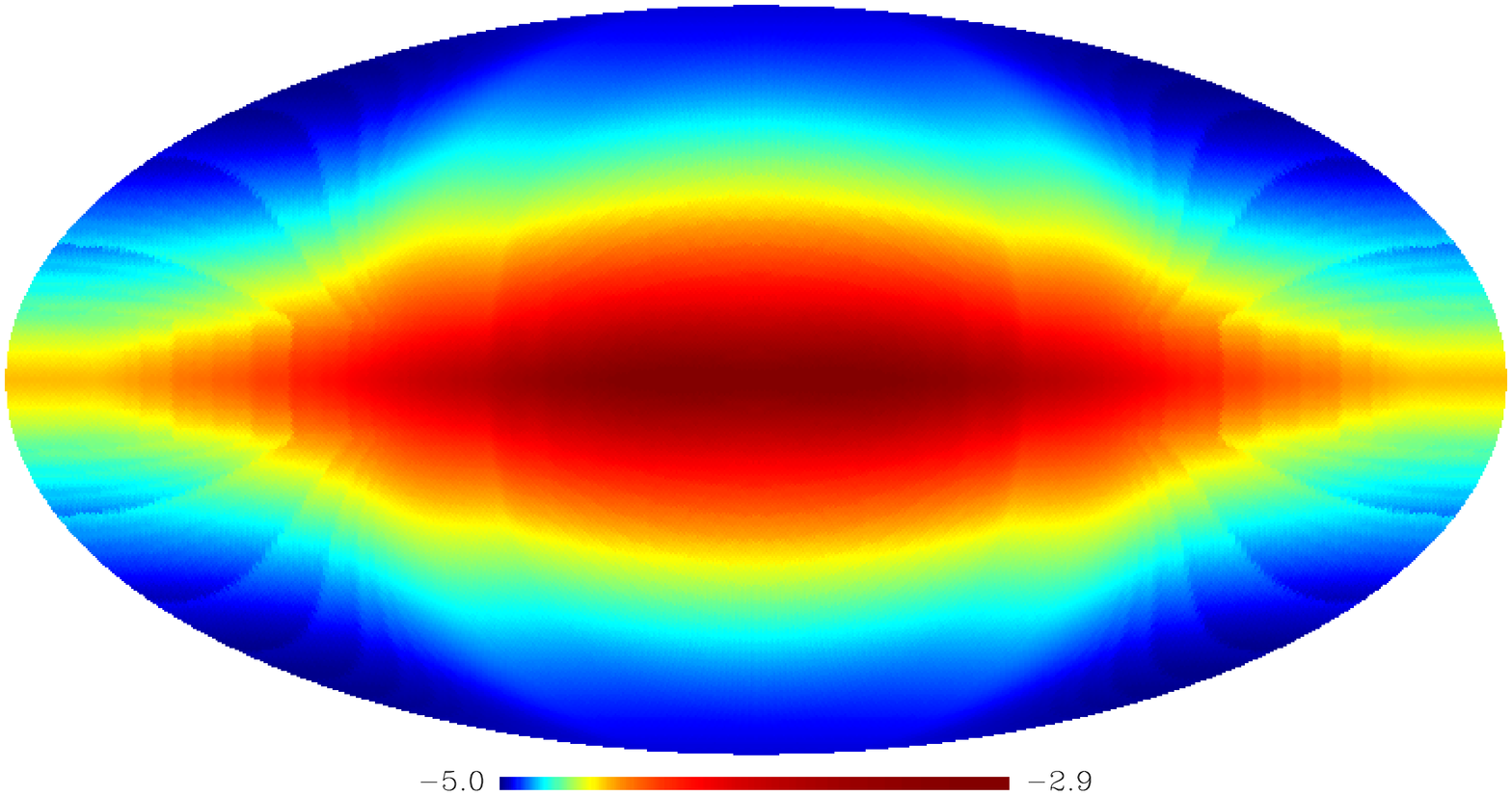}
 \end{minipage}
 \ \hspace{30mm} \
 \begin{minipage}[htb]{5.5cm}
   \centering
   \includegraphics[width=5.5cm,angle=90]{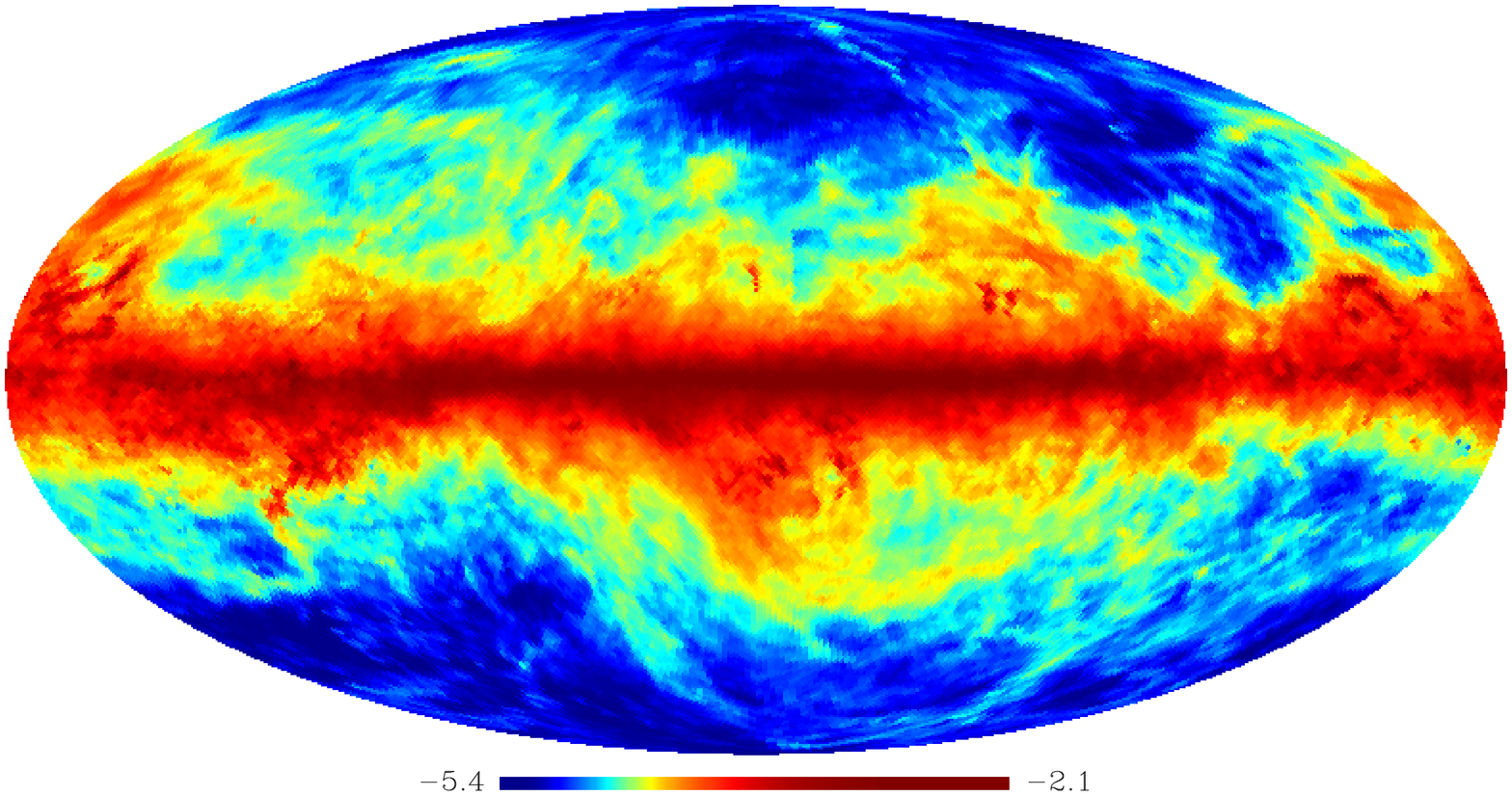}
 \end{minipage}
    \caption{Sky-map at 150 GeV of the emissions associated to Galactic primary+secondary CRs in the "conventional" model B0. The intensity is shown in logarithmic scale and units [MeV cm$^{-2}$ s$^{-1}$ sr$^{-1}$]. {\it Left Panel}: Inverse Compton radiation. {\it Right Panel}: $\pi^0$-decay emission.  }
\label{fig:mapCR}
 \end{figure}

\begin{figure}[t]
 \ \hspace{-35mm} \
 \begin{minipage}[htb]{5.5cm}
   \centering
   \includegraphics[width=5.5cm,angle=90]{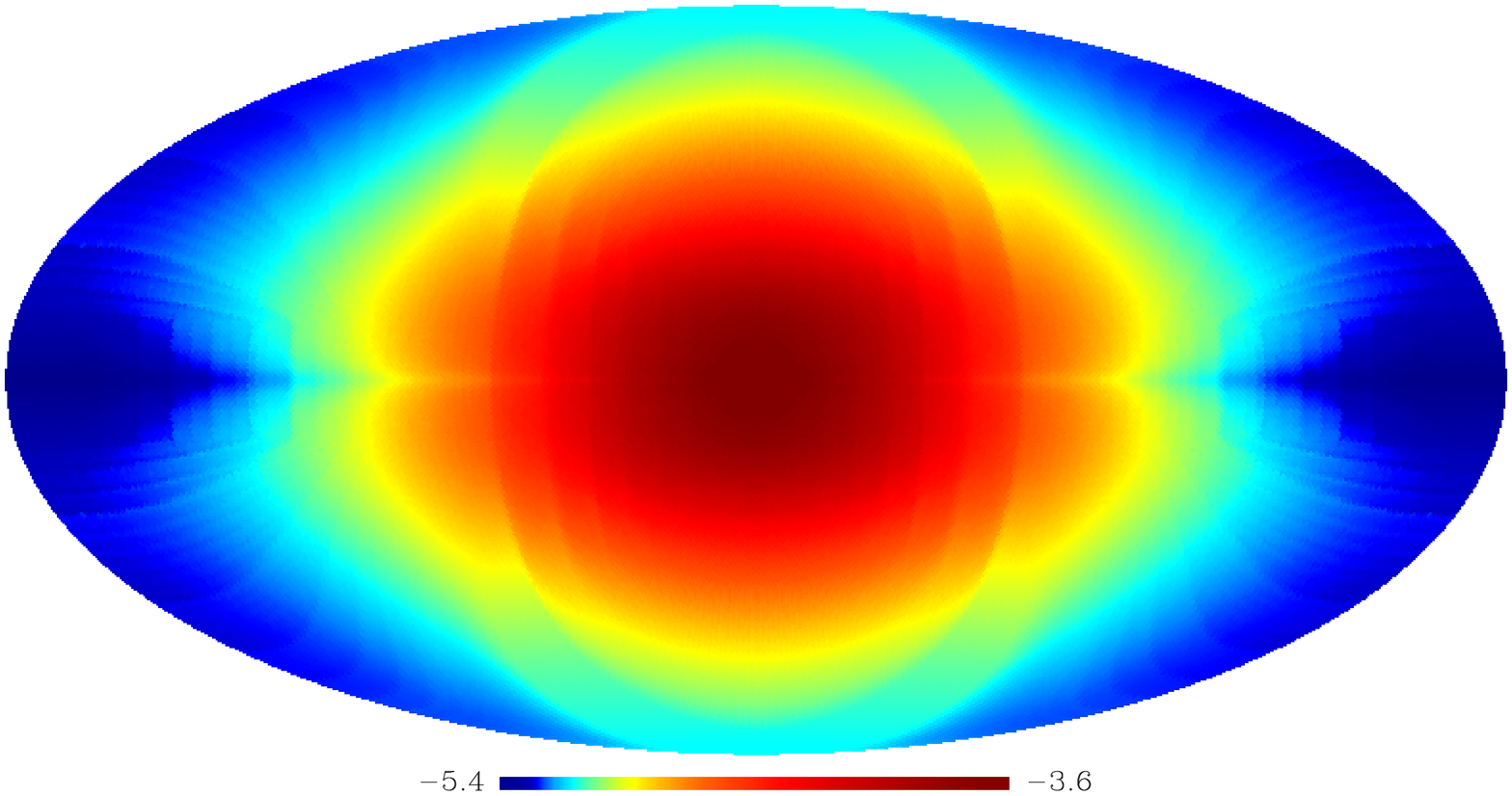}
 \end{minipage}
 \ \hspace{30mm} \
 \begin{minipage}[htb]{5.5cm}
   \centering
   \includegraphics[width=5.5cm,angle=90]{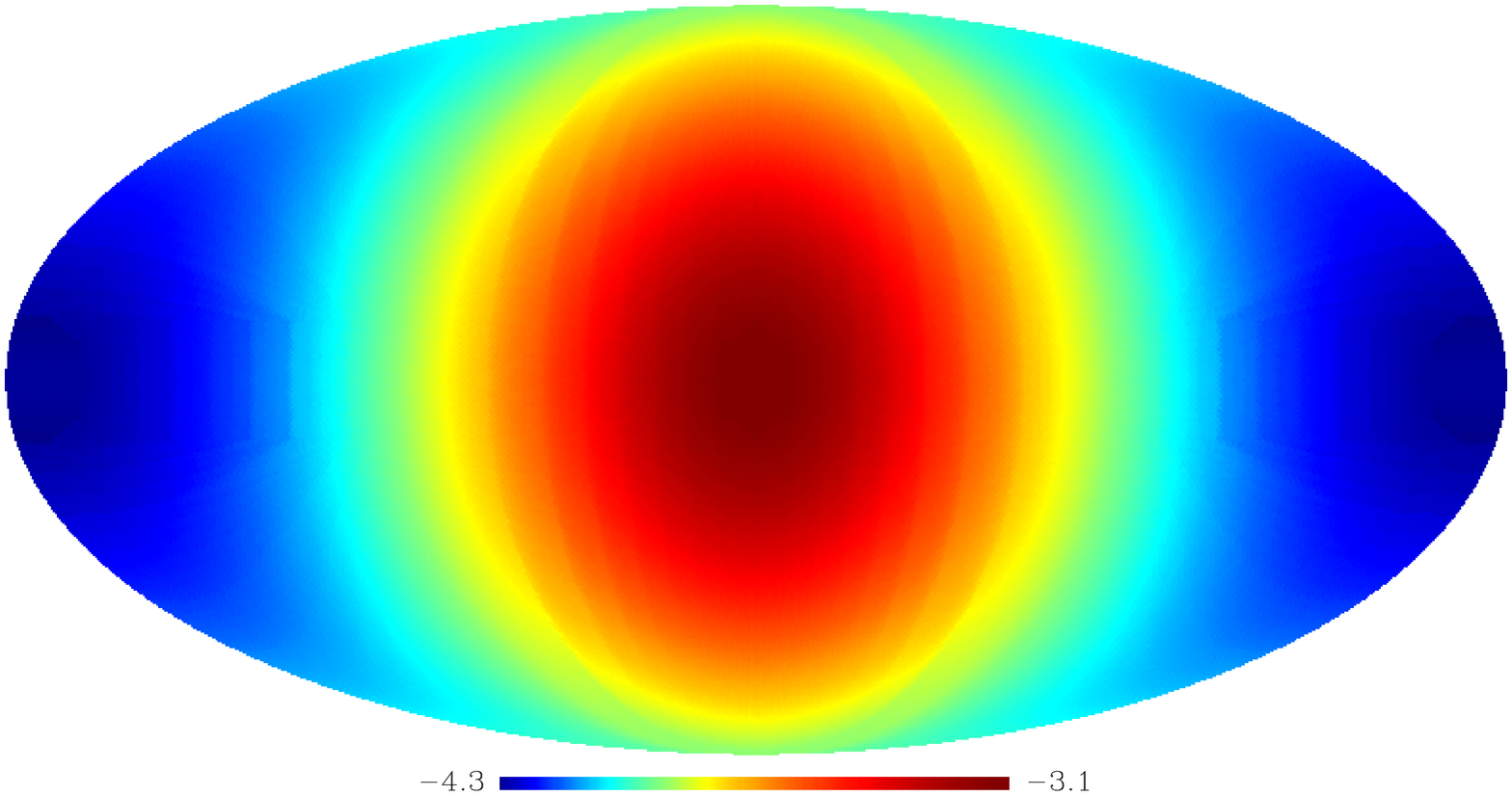}
 \end{minipage}
    \caption{Sky-map at 150 GeV of the emissions induced by WIMP annihilations in the propagation model B0. The intensity is shown in logarithmic scale and units [MeV cm$^{-2}$ s$^{-1}$ sr$^{-1}$]. {\it Left Panel}: Inverse Compton radiation in the DM$\mu$ scenario. {\it Right Panel}: $\pi^0$-decay emission in the DM$\tau$ scenario.}
\label{fig:mapDM}
 \end{figure}

The level we predict for diffuse $\gamma$-ray fluxes is about $E^2 J\simeq 1-3 \cdot 10^{-4}$ MeV cm$^{-2}$ s$^{-1}$ sr$^{-1}$ at $E\gtrsim 100$ GeV (see Figs.~\ref{fig:emissint}-\ref{fig:lonprofhigh}). Considering the FERMI performances stated in Ref.~\cite{Atwood:2009ez} (roughly, an effective area of $A_{eff}=8\cdot 10^3 \, {\rm cm}^2$ and a field of view $FoV=2.4$~sr), the expected number of counts,
for an energy bin size of $\Delta E_{\gamma}=50$~GeV, is about 
$N_{\gamma}\geq 70$ sr$^{-1}$ yr$^{-1}$ . We deduce that the diffuse $\gamma$-ray spectra as predicted in Figs.~\ref{fig:emissint} and \ref{fig:emisshigh} can be detected with a statistical error smaller than $10\%$ in 1 year of observation. The precise description of longitudinal and latitudinal profiles requires, on the other hand, some years of observations. Combining different slices of the sky, however, the disentanglement between the CR source having a "disc" shape and the WIMP induced source having a spherical shape will be feasible in the forthcoming future. Full sky-maps, at 150 GeV for the $\pi^0$-decay signal associated to primary CR and DM$\tau$, and for the IC emission associated to CR electrons and DM$\mu$ is shown in Figs.~\ref{fig:mapCR} and \ref{fig:mapDM}. Differences in morphologies for the various components are indeed very clear.

\subsection{Radio and infrared emission}

Now we turn the discussion on the synchrotron emission in the radio and infrared bands. Electrons and positrons injected by DM or CR source interact with the Galactic magnetic field, giving raise to a synchrotron radiation.
Due to the spectral behaviour, the synchrotron emission is the dominant component of the Galactic diffuse emission at low frequency. The sky-map of Ref.~\cite{Haslam:1982} at 408 MHz is the standard calibration for the synchrotron diffuse signal (although it could include a significant amount of unresolved sources).
Foreground estimations in the WMAP data~\cite{Gold:2008kp} suggest a spectral index for the synchrotron emission $\sim 3$, at frequency up to 60 GHz.
(An anomalous component has be claimed to be present in the innermost region of the Galaxy, a result which depends on the template used for the foreground estimation. The associated spectral index turns out to be harder than 3. Such component, dubbed "WMAP haze", has been associated to be a possible DM signal due to WIMP annihilations~\cite{Finkbeiner:2004us,Hooper:2007kb,Dobler:2007wv,Cumberbatch:2009ji}. Since the haze is associated to the central portion of the Galaxy,
we will not discuss it here.)

\begin{figure}[t]
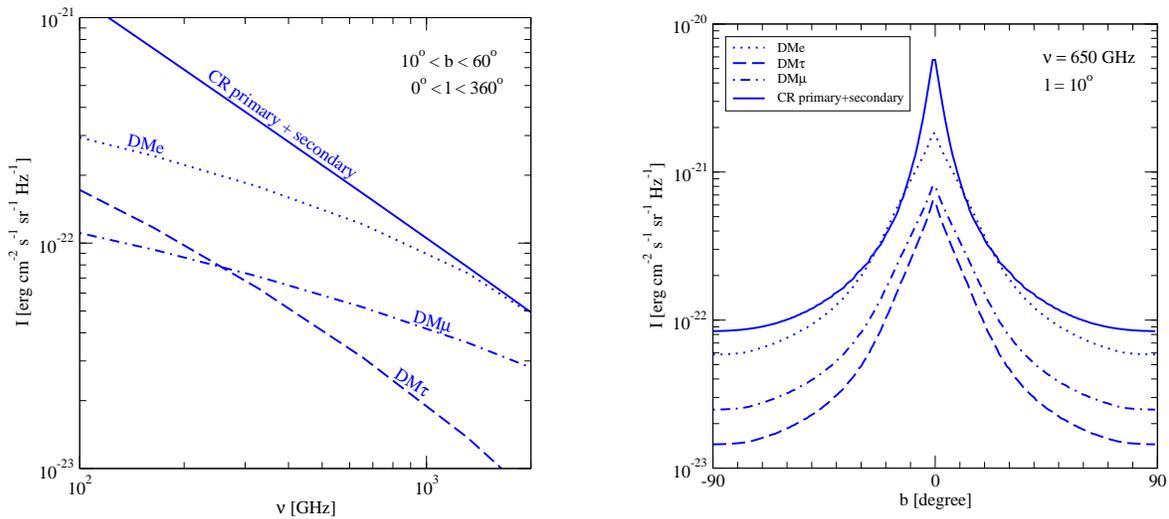

 \begin{minipage}[htb]{7cm}
   \centering
   \includegraphics[width=7.cm]{fig11a.eps}
 \end{minipage}
 \ \hspace{10mm} \
 \begin{minipage}[htb]{7cm}
   \centering
   \includegraphics[width=7.cm]{fig11b.eps}
 \end{minipage}
    \caption{Synchrotron radio and infrared emissions for CR primary + secondary electrons in the "conventional" model B0 (solid) and for the benchmark WIMP scenarios DM$e$ (dotted), DM$\tau$ (dashed), and DM$\mu$ (dashed-dotted). {\it Left Panel}: Spectrum at mid-high latitudes ($10^{\circ}<b<60^{\circ}$) integrated over longitudes $0^{\circ}<l<360^{\circ}$. {\it Right Panel}: Latitudinal profile at "small" longitude ($l=10^{\circ}$) and at frequency $\nu=650$ GHz.}
\label{fig:radio}
 \end{figure}

In Fig.~\ref{fig:radio}, we show the emission associated to primary+secondary CR electrons in the "conventional" model at intermediate latitudes. Matching the diffuse emission induced by CRs with the observed synchrotron emission in the whole Galaxy is beyond the goal of this paper. Note, however, that the spectral index is very close to 3, as required. The overall normalization is also very close to the one estimated by the WMAP team.

Again, in order to explore a possible DM signal, the region at intermediate and large latitudes is the best target. Indeed, the magnetic field slowly decreases outside the disc (we adopt the benchmark case $B=5\,\exp[-(R-R_0)/10\,{\rm kpc}-|z|/2\,{\rm kpc}]\,\mu$G, as described in Section~\ref{sec:CRprop}), and, thus, the signal induced by a spherical DM profile is non-negligible at intermediate and large $z$.

The peak of the synchrotron emissivity is approximately at the frequency
$\nu_p \simeq 4.7 \,{\rm MHz} \,\cdot (E_e/1~{\rm GeV})^2 \, (B/1~\mu{\rm G})$, with $E_e$ the electron energy and $B$ the ambient magnetic field. The injection of high energetic electrons ($E_e\gtrsim 100$~GeV) induce a signal peaked at frequencies $\gtrsim 100$ GHz. It follows that this represents the best frequency range for the search of a synchrotron emission induced by a DM candidate injecting an electron/positron yield at the level of the PAMELA excess. Such frequencies are above the WMAP range, while it could be possible to search for such a component with forthcoming PLANCK satellite, 
which will have detection channels up to a frequency of $850$ GHz.
We show in Fig.~\ref{fig:radio}, the signals induced by the benchmark DM candidates introduced 
above, as computed in the propagation model B0.
The detection prospects are less favourable than in the $\gamma$-ray band. Only the benchmark scenario DM$e$ can induce a significant component at intermediate latitudes.

Moreover, an extra uncertainty in the analysis at infrared frequencies is given by the fact that thermal emission rather than synchrotron is expected to dominate the foreground. Focusing the analysis on the spatial distribution and on polarization data could, however, help to disentangle a DM-induced synchrotron signal.
A firmer statement on such a possibility with the PLANCK satellite deserves a more detailed 
dedicated study.

In principle, data in the X-ray frequency band can also be used to test the diffuse continuum emission induced 
by DM pair annihilations or decays.
On the other hand, the fluxes detected by INTEGRAL SPI~\cite{Bouchet:2008rp} and COMPTEL~\cite{Strong:1998ck} in the inner region of the Galaxy is a factor of few  above the estimated CR diffuse emission (as computed in the "conventional" model). 
It follows that, generically, DM-induced fluxes are in turn less constrained by these data-sets with respect to the $\gamma$-ray and radio observations.
Moreover, the case of a DM candidates which offers an interpretation of the PAMELA excess is even harder to be detected, since the hard injection spectrum of the DM-induced $e^+/e^-$ leads to a signal associated to the IC scattering with the CMB component which is subdominant with respect to the CR contribution, having the latter a softer spectrum (see Figs~\ref{fig:emissint} and \ref{fig:emisshigh}).

\section{Discussion and conclusions}
\label{sec:conc}

We have presented a study of the diffuse $\gamma$-ray emission induced by particle-DM models 
which can account for the PAMELA positron excess. The analysis has been focused on signals induced by pair annihilations (or decays) associated to the smooth component of the Milky Way 
DM halo, in a configuration with no significant enhancement in the DM density in the 
Galactic center region, hence with none of the DM signals usually discussed from the center of the
Galaxy at a detectable level. The goal was to present an analysis based on minimal sets  of
assumptions and extrapolations with respect to locally testable or measurable quantities
(for comparison, e.g., Ref.~\cite{Bertone:2008xr} considers also DM profiles with a central cusp, 
finding strong, though heavily model-dependent, constraints from GC observations).
In this respect, e.g., we are not directly comparing to the possibility of detecting DM fluxes from 
Milky Way satellites or other dark matter dominated objects, avoiding the extrapolations that are
required when addressing this kind of predictions. Estimates of extragalactic $\gamma$-ray 
background from unresolved DM structures lead typically to a flux below the astrophysical 
EGB (unless a substantial enhancement stems from populations of dense substructures), 
which is mainly related to unresolved blazars; this component is therefore marginally relevant for 
our analysis (although it could be possibly detected studying its anisotropy pattern~\cite{Ando:2006cr}).
On the other hand, a contribution of unresolved Galactic subhalos to the diffuse emission of the 
Galaxy could be substantial~\cite{Pieri:2007ir}. Our conclusions would be strengthened, in particular, 
in case of a spatial distribution of subhalos antibiased with respect to the host halo mass 
distribution~\cite{Kuhlen:2008aw}, which would give a larger diffuse emission at high latitudes and longitudes. While not changing the general picture, the estimation of this component depends on 
several quantities which are not well-known, such as the mass function, the spatial distribution and 
the concentration of subhalos.

The presence of a Galactic dark disc can in principle affect our conclusions if the DM density associated to the disc is much higher than the density of the halo. However, this is quite unlikely, as also found in the simulation of Ref.~\cite{Read:2008fh}, which proposed the existence of a dark disc in Milky Way-sized galaxies.
Moreover, we consider, for simplicity, a spherical shape for the DM halo, which is actually an approximation.
Significant deviations from sphericity could induce a different scaling of the signal in the longitudinal direction with respect to the latitudinal direction. However, our general conclusions will be unchanged, being crucially based only on the assumption that the DM halo is extended along the z-direction well further the stellar disc.

We have compared the DM signal to standard astrophysical components, analyzing spectral and 
spatial features to derive which are the best energy band and region of the sky to disentangle the DM contribution from the Galactic plus extragalactic background with the FERMI LAT telescope.
Being the sources of acceleration for CR (namely, Supernovae Type II) distributed within a thin disc, 
the ratio of the associated background versus the DM signal decreases as the latitude increases.
The extragalactic background has, on the other hand, a flat spatial distribution, and dominates at very large latitudes. Intermediate and moderately large latitudes ($10^{\circ}\leq b \leq 60^{\circ}$) are thus the most promising portion of the sky for the search of a diffuse emission induced by Galactic DM. 

Fits to the PAMELA positron excess require sources with a rather hard spectrum. The preferred 
DM-related interpretation is a scenario with prompt emission of  leptonic final states; we have 
discussed in detail three sample benchmark cases, namely emission of monochromatic $e^+/e^-$, 
the case of $\tau^+/\tau^-$ yields, giving a shallower $e^+/e^-$ spectrum, and the channel $\mu^+/\mu^-$ as final state of annihilation/decay. 
We have disregarded WIMP models with annihilation channels producing soft spectra, 
like quark-antiquark pairs, and with weak gauge boson final states, which would overproduce cosmic-ray antiprotons~\cite{Donato:2008jk}.

We find that for all the benchmark models, after tuning the source rate to match the level of the PAMELA positron excess, the DM-induced $\gamma$-ray flux becomes comparable to the background 
at energies $E\gtrsim 100$ GeV. In case of pairs annihilation into monochromatic $e^+/e^-$ or into $\mu^+/\mu^-$, 
$\gamma$-rays arises from IC scattering of the propagating $e^+/e^-$ and FSR processes 
at emission; the scenario is detectable for sufficiently heavy DM candidates 
($M_\chi \gtrsim 150$~GeV; the PAMELA positron data require $M_\chi \geq 80$~GeV). In the case 
of a DM candidate annihilating into $\tau^+ \tau^-$ with a mass of few hundreds of GeV, as 
required by the PAMELA excess without violating ACT constraints, the detectable 
$\gamma$-ray component is due to the emission from $\pi^0$-decays, which peaks at, roughly, 
one-third of the DM mass. These conclusions can be straightforwardly extended to a decaying DM scenario: the mass scale just needs to be doubled and the decay rate properly adjusted.
In case of a DM candidate annihilating into a new light particle which in turn decays into leptons, the FSR is generally reduced (depending on the model). The total emission can be thus mildly reduced with respect to an analogous WIMP case, but remains still sizable. 

The statement regarding  the detectability of the induced Galactic diffuse $\gamma$-ray flux has
a marginal dependence on the model implemented to describe the propagation of charged 
particles in the Galaxy. We introduce a locally self-consistent picture, implementing for 
$e^+/e^-$ (as well as for primary CR protons and all other relevant components) a transport 
equation with propagation parameters tuned to reproduce the LIS of cosmic rays. 
In particular, being mostly sensitive to the picture at GeV to TeV energies, we fit propagation
parameters to B/C data at $E\geq 3$ GeV. We consider as a reference model the "conventional" 
model introduced by the GALPROP collaboration~\cite{Strong:1998pw}. We then discuss the 
variation of the vertical profiles of CR and DM-induced electrons and of the associated diffuse 
emissions by changing the halo boundaries of propagation, introducing an advection term, 
modifying the spectral index for diffusion, and considering a spatially varying diffusion coefficient.
We have found that the signal to background ratio is enhanced if the halo boundaries in the $z$-direction are extended with respect to the "conventional" case ($z_h=4$ kpc), and reduced 
if they are restricted. The latter possibility is, however, disfavoured by CR data. 
For all the other propagation models, the detection prospects are analogous to the 
"conventional" case.

We have also discussed the possibility of detecting radiative emissions at other wavelengths. 
In the radio and infrared bands, the synchrotron component associated to a scenario in which 
DM injects a very hard $e^++e^-$ spectrum (like, e.g., a WIMP with a large branching ratio of annihilation into $e^+/e^-$) is expected to be a significant component of the synchrotron diffuse emission of the 
Galaxy at few hundreds of GHz. This frequency range can be probed by the forthcoming 
PLANCK mission.
On the other hand, this picture is strongly constrained by the FERMI data on $e^++e^-$, and the detection prospects at radio frequencies are, typically, less favourable than in the $\gamma$-ray band. 
For the class of DM candidates explaining the raise in the positron fraction, 
the X-ray energy band is even less favourable, being the IC spectrum induced 
by DM harder than the one associated to CRs. 

To conclude, we have found that the DM interpretation of the PAMELA positron excess can be 
tested by the FERMI LAT telescope in the diffuse emission at mid-latitudes and high energy.
In the case of a DM candidate injecting a very hard $e^+/e^-$ yield, the signal is given by the 
IC + FSR emission. The case of a DM candidate with a sizable and hard $\gamma$-ray yield 
from  $\pi^0$-decay (like the case of $\tau^+/\tau^-$ as final state of annihilation/decay), can be 
even more easily tested. A crucial ingredient for the discussion is the different spatial profile for 
the CR primary sources (confined to the disc) and of dark matter induced components 
(spherical distribution). The two terms can be disentangled looking at the angular profile
of the diffuse emission, with the optimal region to single out the DM component being at
intermediate latitudes. Such cross-correlation test of the PAMELA excess would then be 
performed by focusing on a nearby portion of the Galaxy, where the extrapolation on the
DM density profile as well as on the propagation model parameters (from the 
locally-measured CR spectra) can be regarded as rather robust. Indeed, would 
FERMI discover an extra $\gamma$-ray term, with the spectral and angular features matching
the features we discussed for the DM source, this would be a solid step towards the identification 
of the DM component of the Universe.

In turn, would FERMI find that the $\gamma$-ray diffusion emission is instead in agreement 
with the prediction from standard CR components only, tight constraints on the DM interpretation 
of the PAMELA positron excess would follow. On the other hand, such picture would not be in 
contradiction with other scenarios addressing the PAMELA excess. If the additional $e^+/e^-$ 
sources have a spatial distribution analogous to the standard CR source distribution, 
such as, e.g., in the case of pulsars or secondary production inside CR sources, it will be very hard 
to single out the associated Galactic diffuse emission. Indeed, it would be, typically,
subdominant and with a signal to background ratio at the same level in any portion of the 
sky; hence, a different strategy would be needed to cross-check these interpretations of the 
PAMELA positron excess.

\section*{Acknowledgements}
M.R. acknowledges funding by, and the facilities of, the Centre for High Performance Computing, Cape Town.

P.U. is partially supported by the European Community's Human Potential Programme under contracts  MRTN-CT-2006-035863.

\end{document}